\documentclass[preprint2]{aastex}

\usepackage{epsf}
\usepackage{graphics}
\usepackage{color}

\newcommand{\be}{\begin{equation}}
\newcommand{\ee}{\end{equation}}

\newcommand{\ax}{$\alpha_{\rm X}$}
\newcommand{\auv}{$\alpha_{\rm UV}$}

\newcommand{\aox}{$\alpha_{\rm ox}$}

\newcommand{\rb}[1]{\raisebox{1.5ex}[-1.5ex]{#1}}
\newcommand{\msun}{$M_{\odot}$}

\newcommand{\plm}{$\pm$}

\newcommand{\swift}{{\it Swift}}
\newcommand{\xmm}{{\it XMM-Newton}}
\newcommand{\chandra}{{\it Chandra}}
\newcommand{\wpvs}{{WPVS~007}}
\newcommand{\lledd}{$L/L_{\rm Edd}$}
\shorttitle{Multi-wavelength variability of WPVS 007}
\shortauthors{Grupe et al.}

\begin{document}

%\input DGrupe_clipfig.tex
%\useunitmm

\def\etal{{\it et\thinspace al.}\ }
\def\alp{{$\alpha$}\ }
\def\al2{{$\alpha^2$}\ }


\title{Strong UV and  X-ray  variability of the Narrow Line 
Seyfert 1 Galaxy WPVS 007 -- on the nature of the X-ray low state 
}


\author{Dirk Grupe\altaffilmark{1},
S. Komossa\altaffilmark{2},
Julia Scharw\"achter\altaffilmark{3,4},
Matthias Dietrich\altaffilmark{5,6},
Karen M. Leighly\altaffilmark{7},
Adrian Lucy\altaffilmark{7},
Brad N. Barlow\altaffilmark{1},
}

\altaffiltext{1}{Department of Astronomy and Astrophysics, Pennsylvania State
University, 525 Davey Lab, University Park, PA 16802; email: dxg35@psu.edu} 

\altaffiltext{2}{Max-Planck-Institut f\"ur Radioastronomie, Auf dem H\"ugel 69,
53121 Bonn, Germany}

\altaffiltext{3}{Research School of Astronomy and Astrophysics, 
The Australian National University, Mount Stromlo Observatory, 
Cotter Road, Weston Creek, ACT 2611 Australia}

\altaffiltext{4}{Observatoire de Paris, LERMA,
61 Avenue de l'Observatoire, F-75014 Paris, FRANCE, email: julia.scharwaechter@obspm.fr}

\altaffiltext{5}{Department of Astronomy, The Ohio State University, 140 West 
18th Avenue, Columbus, OH 43 210, USA}

\altaffiltext{6}{Dept. of Physics and Astronomy, Clippinger Labs 251B, Athens, OH 45701}

\altaffiltext{7}{Homer L. Dodge
Department of Physics and Astronomy, University of Oklahoma, 
440 West Brooks Street, Norman, OK 73019; email: leighly@nhn.ou.edu}


%\altaffiltext{1}{Visiting Astronomer, Cerro Tololo Inter-American Observat}


\begin{abstract}
We report on multi-wavelength observations of the X-ray transient 
Narrow Line Seyfert 1 (NLS1) galaxy
WPVS 007. The galaxy was monitored with \swift\ between October
2005 and July 2013, after it had undergone a dramatic
drop in its X-ray flux earlier. 
For the first time, we are able to repeatedly detect this
NLS1 in X-rays again. This increased number of
detections in the last couple of years may suggest that the strong absorber that
has been found in this AGN is starting to become leaky, and may eventually disappear.
The X-ray spectra obtained for \wpvs\ are all consistent with a partial
covering absorber model. 
A spectrum based on the data 
 during the extreme low X-ray flux states shows that the
absorption column density is of the order of $4\times 10^{23}$ cm$^{-2}$ 
with a covering fraction of 95\%.
WPVS 007 also displays one of the strongest UV variabilities seen in 
Narrow Line Seyfert 1s. The UV continuum
variability anti-correlates with the optical/UV slope \auv\ 
which suggests
that the variability primarily may be due to reddening. 
The UV variability time scales are consistent with 
moving dust `clouds' located beyond the dust sublimation 
radius of $R_{\rm sub} \approx$ 20 ld. 
We present for the first time near infrared JHK data
of WPVS 007, which reveal a rich emission-line spectrum.
Recent optical spectroscopy does not indicate significant variability
in the broad and FeII emission lines,  implying that
the ionizing continuum seen by those gas clouds
has not significantly changed over the last decades.
 All X-ray and UV observations are consistent with a scenario
in which an evolving Broad Absorption Line (BAL) flow obscures the continuum emission. 
  As such, WPVS 007
is an important target for our understanding of BAL flows in low-mass
active galactic nuclei (AGN). 
\end{abstract}


\section{Introduction}

Outflows are a ubiquitous property of AGN. 
For example, blue-shifted emission lines 
like [OIII]$\lambda$5007 \citep[e.g., ][]{komossa08},
or blue shifted absorption lines in the UV and in X-rays \citep[e.g., ][]{crenshaw04} are typically
interpreted as signs of outflowing gas. 
Outflows can be driven in principle magnetically, thermally, and through radiation
\citep[e.g.,][]{kurasawa09a,kurasawa09b,proga04}.
With the kinetic energy and angular momentum transported outwards, outflows have 
strong influences on the AGN environment. As a consequence, many AGN parameters  
are driven by 
the Eddington ratio \lledd\ \citep[e.g., ][]{boroson02, sulentic00, grupe04, xu12}.
 The class of AGN that shows 
 the strongest outflows (besides jets) are Broad Absorption
Line Quasars \citep[BAL QSOs, e.g.,][]{weymann91} 
which can reach outflow velocities of 
 more than 20000 km s$^{-1}$ \citep[e.g.,][]{hamann08}.
Roughly 10-20\% of optically-selected 
quasars belong to this class \citep[e.g.,][]{dai08, elvis00} with increasing percentage
among infrared selected samples, 
which is often interpreted as
an inclination effect.
However, it has also been suggested that the occurrence
of BALs may mark a specific time in the life of a quasar \citep{mathur00, becker00}.
Because the strength of radiation driven outflows directly
depends on \lledd, BAL QSOs are at one extreme end of the \lledd\
 distribution \citep[e.g.,][]{boroson02}. In the local Universe the AGN with
 the highest \lledd\ are Narrow Line Seyfert 1 galaxies \citep[NLS1s;][]{osterbrock85}. 

NLS1s have drawn a lot of attention over the last two decades due to their extreme 
properties, such as, on average, steep X-ray spectra, strong Fe II emission and weak
emission from the Narrow Line Region \citep[e.g.,][]{boroson92, grupe04, komossa08r}. 
All these properties are linked, 
and are most likely driven primarily by the mass of the central black hole and 
the Eddington ratio \lledd.  Generally speaking, NLS1s are  AGN with 
low black hole masses and 
high \lledd.
NLS1s have also been considered to be AGN in an
early stage of their evolution \citep[][]{grupe99, mathur00}.

It has been suggested by \citet{brandt00b} and \citet{boroson02}
that BAL QSOs and NLS1s are similar with respect to their 
high \lledd. However, they differ in their black hole masses and 
appear to be at opposite ends of the $M_{\rm BH}$
spectrum. It has also been found that the rest-frame
optical spectra of at least some
BAL QSOs look very much like low-redshift NLS1s 
\citep[e.g.,][]{marziani09, dietrich09}.
While all these results seem to
support the connection between BAL QSO and NLS1s, what is missing is a direct
link between them: an AGN that shows properties of a typical NLS1 as well as
those of BAL QSOs. Such a 
link is the NLS1 WPVS 007.

When the NLS1 \wpvs\ 
\citep[RBS 0088; 1RXS J003916.6$-511701$;
$\alpha_{2000}$ = $00^{\rm h} 39^{\rm m} 15.^{\rm s}8$, 
$\delta_{2000}$ = $-51^{\circ} 17' 03 \farcs 0$, z=0.02861; ][]{wamsteker85,grupe95,schwope00}
was discovered in X-rays during the  ROSAT All Sky 
Survey \citep[RASS;][]{voges99} it appeared to be a
normal X-ray bright AGN, although its X-ray
   spectrum was unusually soft \citep{grupe95}.
However, when it was observed again three years later with the ROSAT Position Sensitive 
Proportional Counter \citep[PSPC; ][]{pfeffermann86} it appeared to have 
almost vanished in
X-rays, with a factor of more than 400 drop in its X-ray flux
 \citep{grupe95}. In the following years it was observed several times by 
ROSAT \citep{grupe95} and Chandra in 2002 \citep{vaughan04}. During all these
observations, WPVS 007 was extremely X-ray faint. 
We started monitoring \wpvs\ with the NASA Gamma-Ray Burst Explorer mission
\swift\ \citep{gehrels04}
in October 2005 \citep{grupe07a} and were able 
to finally detect it with \swift\ in a 50 ks observation in September 2007 \citep{grupe08b}.
This observation led to the detection of hard X-ray photons in the X-ray spectrum
of \wpvs\ for the first time. The hard X-ray spectrum suggested the presence of a strong 
partial covering absorber. 

The strongest clue that the X-ray transience is due to absorption, however, 
came from UV spectroscopy.
In 1996, WPVS 007 was observed by HST \citep{goodrich00}
and low-velocity broad absorption troughs (mini BALs)
were present in the UV resonance 
lines. As suggested by \citet{hamann08} mini BALs may be the early (or late)
states of BALs. When
WPVS 007 was observed again in the UV by FUSE in 2003 
it had developed a very strong
broad absorption line flow \citep{leighly09}. 
HST COS UV spectra obtained in 2010 June
showed a dramatic increase in the UV SiIV and CIV
absorption lines \citep{cooper13} compared with the 
1996 HST observation.

\wpvs\ is special because BALs are rare in such low-luminosity
systems as WPVS 007 (e.g., \citep{laor02}, and because it allows us to 
study properties which otherwise can not be studied easily 
in BAL QSOs. Due to their large black hole masses, the time scales
observed in BAL QSOs are very long, and at least on the order of years 
\citep[e.g.,][]{capellupo11, saez12}.
In \wpvs, however, time scales are shorter, as it is ~100 times
less luminous than a typical BALQSO and with a similarly smaller black
hole mass and typical size scales
 \citep{leighly09}.

The outline of this paper is as follows: in \S\,\ref{observe} we describe the
observations by \swift, in the near infrared, and in the optical, as well as the 
data reduction.  In \S\,\ref{results} we 
present the results from the analysis of the light curves and 
X-ray spectroscopy of the \swift\ XRT data and the optical and near infrared spectroscopy.
In \S\,\ref{discuss} we discuss the results. 
Throughout the paper spectral indices are denoted as energy spectral indices
with
$F_{\nu} \propto \nu^{-\alpha}$. Luminosities are calculated assuming a $\Lambda$CDM
cosmology with $\Omega_{\rm M}$=0.27, $\Omega_{\Lambda}$=0.73 and a Hubble
constant of $H_0$=75 km s$^{-1}$ Mpc$^{-1}$. This results in a luminosity distances $D$=118 Mpc
using the cosmology calculator  by \citet{wright06}. All errors are 1$\sigma$ unless stated otherwise.

\section{\label{observe} Observations and data reduction}

\subsection{\swift\ Observations}

Table\,\ref{swift_log} presents the 
\swift\ observations of \wpvs\ starting on 2008
January 15, including the start and end times and the total exposure
times. The lists of the previous  \swift\ observations between 2005
October and 2007 December can be found in \citet{grupe07a} and \citet{grupe08b}.  
The \swift\ X-ray telescope \citep[XRT;][]{burrows05} 
was operating in photon counting mode \citep{hill04} and the
data were reduced by the task {\it xrtpipeline} version 0.12.6., 
which is included in the HEASOFT package 6.12. Source counts were selected in a
circle with a radius of 24.8$^{''}$ and background counts in a nearby 
circular region with a radius of 247.5$^{''}$. The 
 3$\sigma$ upper limits and the
count rates of the detections were determined by applying the Bayesian method by \citet{kraft91}.
For these detections, due to the low number of counts, we applied Bayesian statistics to
determine the hardness ratios as described by \citet{park06}.
Some of the detections allowed a spectral analysis using Cash statistics \citep{cash79}. 
In some cases, like in the already published observation from September 2007, the data from 
several segments when \wpvs\ was detected were merged. For this purpose we created auxiliary response files (ARF)
 for each segment
and then combined these into one arf by using the FTOOL {\it addarf}. 
For all spectra we used the most recent response file {\it swxpc0to12s6\_20010101v013.rmf}.
Due to the small number of counts used in the spectra, the counts were not binned.

The UV-optical telescope \citep[UVOT;][]{roming05} 
data of each segment were coadded in each filter with the UVOT
task {\it uvotimsum}.
Source counts in all 6 UVOT filters
  were selected in a circle with a radius of 5$^{''}$ and background counts in
  a nearby source free region with a radius of 20$^{''}$.
  UVOT magnitudes and fluxes were measured with the task {\it  
uvotsource} based on the most recent UVOT calibration as described in  
\citet{poole08} and  \citet{breeveld10}.
The UVOT data were corrected for Galactic reddening
\citep[$E_{\rm B-V}=0.012$; ][]{sfd98}. The correction factor in each  
filter was
calculated with equation (2) in \citet{roming09}
who used the standard reddening correction curves by \citet{cardelli89}.
Due to new UVOT calibration files \citep{breeveld10} also observations that were previously published in
 \citet{grupe07a} and \citet{grupe08b} were reanalyzed in order to have a consistent data set. All these 
 UVOT data starting in October 2005 are listed with their magnitudes corrected for Galactic reddening 
 and the optical/UV spectral slope \auv\ in Table\,\ref{swift_results}. This optical/UV spectral
 slope \auv\ was determined from a power law fit to the UVOT data. 
Note that although we generally observed \wpvs\ with the \swift\ UVOT in all 6
filters since the beginning of the monitoring campaign in 2005, we switched to
the uv m2 filter only after the end of the \swift\ guest investigator program in
March 2012 in order to save on UVOT filter wheel rotations. The uv m2 filter was
picked because it is the cleanest UV filter with no broad wings in the filter 
response \citep{breeveld10}.

In order to obtain spectral energy distributions, a source and background spectral file 
was created for each filter
using the UVOT task {\it uvot2pha}.  Also here the latest UVOT response files for each filter were used. 

\subsubsection{Near Infrared (NIR) Observations}

We observed \wpvs\ as part of an observing program using the NIR imaging 
spectrograph SofI at the 3.5 m NTT at
 La Silla/ESO, to study the rest-frame 
optical spectra of a sample of BAL QSOs \citep{dietrich09}. The observation was performed on 
2004 September 12 (MJD 53260.112)
for $18\times 180$s  in the J plus H,  and $8\times 180$s in the H plus K band wavelength ranges
 under photometric conditions with 1$^{''}$ seeing. A 1$^{''}$ long slit was 
used for the observation. We used an $1024\times1024$ pixel$^{2}$ Rockwell HAWAII HgCdTe detector. 
The data were reduced with ESO's Munich Imaging and Data Analysis System, MIDAS. 
The data reduction was carried out the same way as in \citet{dietrich09},
and a full description 
can be found in that work. 

\subsubsection{Optical Spectroscopy}
An optical spectrum of WPVS 007 was obtained on 30 September 2011 using the Wide 
Field Spectrograph WiFeS \citep{Dopita07, Dopita10} at the ANU 2.3m telescope at 
Siding Spring Observatory (Australia). WiFeS is an integral field spectrograph with a 
field of view of $38\arcsec \times 25\arcsec$ produced by 25 slitlets of 
$38\arcsec \times 1\arcsec$. The instrument provides a wide wavelength coverage via 
simultaneous observations in a blue and red arm at spectral resolutions of 3000 or 7000. 
WPVS 007 was observed using the B3000 grating in the blue arm and the R7000 grating in 
the red arm. The data presented here are the median average of five exposures of 600~s 
each. The observations were performed in nod-and-shuffle mode for sky-subtraction,
using 100~s integrations on the object and the sky. A bias frame and a Ne-Ar lamp 
exposure were taken between the object exposures in order to be able to account for 
variations in the bias level and the wavelength solution during the night. The data 
reduction is based on the reduction recipes implemented in the IRAF reduction package 
for WiFeS \citep[see][]{Dopita10}. A B4V star, observed shortly before WPVS 007, is used 
as a template star for correcting telluric features in the red part of the spectrum. 
The spectrophotometric calibration is based on observations of CD-30~18140 obtained 
during the same observing night. Since the night was partly cloudy, the absolute flux 
calibration is uncertain. In this 
paper, we present the integrated nuclear spectrum of \wpvs\ obtained for 
a spatial aperture of 1 arcsec.

On 23 October 2012, we obtained multiple spectra of \wpvs\ with a total integration time of 1800 s 
using the Goodman High Throughput Spectrograph on the 4.1-m SOuthern Astrophysical Research 
(SOAR) telescope \citep{clemens04}. 
  We used the 0".84  long slit in conjunction with the 930 1/mm VPH grating, which has a dispersion
   of 0.417 Å per unbinned pixel.  This setup allowed us to cover the spectral 
   range 4130-5830 \AA with a 
   resolution of 2.3 \AA\ (R=2100); the mean signal-to-noise ratio in the combined spectrum was approximately 
   110 per resolution element.  Standard IRAF routines were used to bias-subtract, flat-field, 
   and wavelength-calibrate (using FeAr lamp spectra) the spectral frames.  
   Flux calibration was achieved using observations of the spectro-photometric standard star LTT 3218 (spectral type: DA) 
   obtained using the same instrumental configuration as \wpvs.

\section{\label{results} Results}

\subsection{X-ray Variability \label{x-detect}}

\begin{figure*}
\epsscale{1.8}
\plotone{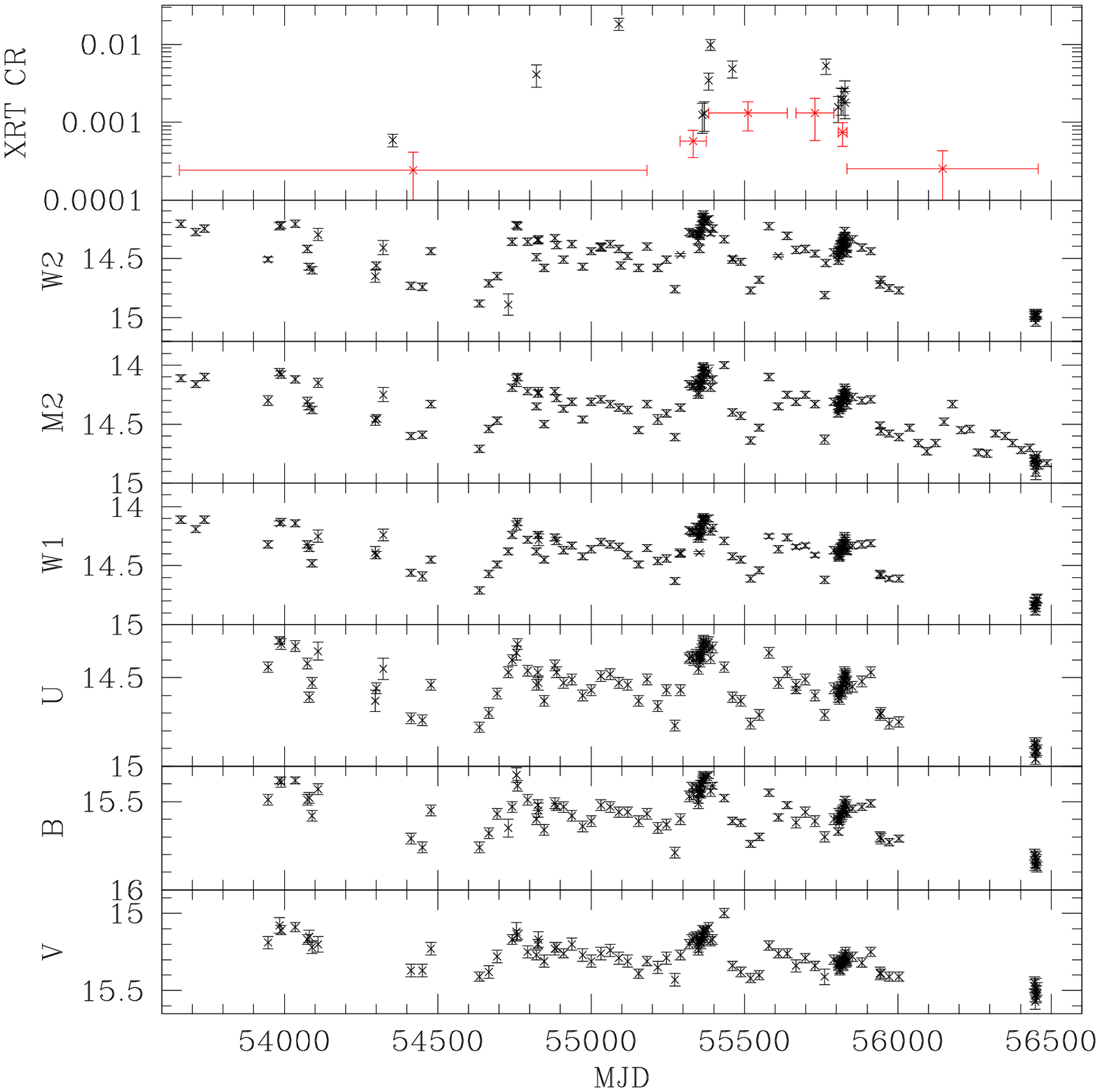}
\caption{\label{wpvs007_xrt_uvot_lc} \swift\ XRT and UVOT light curves of WPVS 007 starting in 2005
October until 2013 July. The X-ray detections of \wpvs\ are listed in Table\,\ref{xrt_stat} and all
reddening corrected UVOT magnitudes are listed in Table\,\ref{swift_results}.
The red data points  mark the long-term detections of \wpvs\ when it was undetected in a single observation. 
Data were added until the source was detected at a 3$\sigma$ level. 
}
\end{figure*}

\begin{figure}
\epsscale{0.6}
\plotone{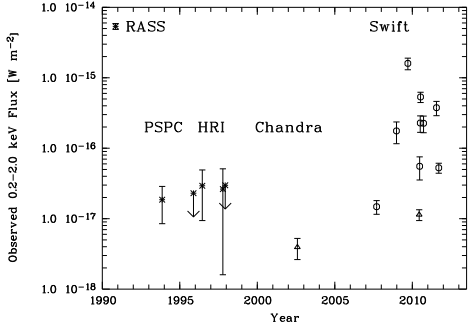}
\caption{\label{wpvs007_xray_lc} Long-term 0.2-2.0 keV X-ray light curve of WPVS 007
 starting with the 
ROSAT All-Sky Survey observation in November 1990. ROSAT observations are marked as stars, Chandra data as triangles,  
\swift\ detections (Table\,\ref{xrt_stat}) as open circles. 
Note that the detection limit of the \swift\ XRT for a 5ks observation is about 4$\times 10^{-17}$ W m$^{-2}$.
}
\end{figure}

Table\,\ref{xrt_stat} lists the \swift\ detections of \wpvs.
The \swift\ X-ray light curve with these detections is displayed in 
Figure\,\ref{wpvs007_xrt_uvot_lc} 
which also shows the light curves in each of the 6 UVOT filters. 
The first detection of WPVS 007 by \swift\ was obtained in 2007 September when \swift\ collected
 more than  50 ks of data on
the source \citep{grupe08b}. The X-ray light curve of the top panel of
Figure\,\ref{wpvs007_xrt_uvot_lc} also shows the count rates determined from
the data when \wpvs\ is not detected in a single observation with the \swift\
XRT. These data points are displayed as red crosses.
The entire long-term X-ray 0.2-2.0 keV flux
light curve starting with the discovery during the RASS is displayed in
Figure\,\ref{wpvs007_xray_lc}. 

The strongest \swift\ detection 
was found on 2009 September 17 when 
\wpvs\ was at a level of $1.6\times 10^{-15}$ W m$^{-2}$ (equivalent to
$1.6\times 10^{-12}$ erg s$^{-1}$ cm$^{-2}$)
which was just a factor of 3.5 fainter
than during the detection during the RASS. This `high state' however, lasted only for a few days.

When we observed \wpvs\ again with \swift\ a week later in a ToO observation,
we did not detect the AGN anymore. The 3$\sigma$ upper limit at the position of \wpvs\ was
$2.3\times 10^{-3}$ counts s$^{-1}$ which converts to $2.0\times 10^{-16}$ W m$^{-2}$ assuming the 
spectrum of the 2009 September 17 (MJD 55091)
observation. 

\begin{figure}
\epsscale{0.6}
\plotone{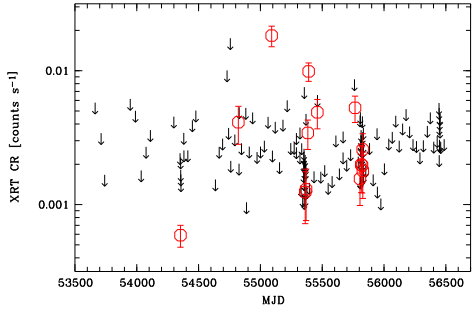}
\caption{\label{wpvs007_xray_ul} \swift\ XRT 3$\sigma$ upper limits (downward arrows) and detections
(open circles) of WPVS 007 as listed in Tables\,\ref{swift_xrt_3sul} and \ref{xrt_stat}, and
Table 1 in \citet{grupe07a}.}
\end{figure}

We note that Figure\,\ref{wpvs007_xray_lc} only displays the X-ray detections 
of \wpvs, except for two upper limits in the ROSAT HRI. 
However,
typically \wpvs\ is not detected by the \swift\ XRT in the 2 or 5 ks 
monitoring observations. The detection limit for
a 5ks \swift\ XRT observation is of the order of $1\times 10^{-3}$ 
counts s$^{-1}$ and the 3$\sigma$ upper limit for a 
2 ks observation is of the order of 3$\times 10^{-3}$ counts s$^{-1}$. 
The \swift\, XRT 3$\sigma$ upper
limits are listed in Table\,\ref{swift_xrt_3sul} and displayed together with the XRT detections in
Figure\,\ref{wpvs007_xray_ul}. This figure clarifies that most of the time \wpvs\ is un-detectable by
\swift\ and that \wpvs\ is typically at a count rate below about 1$\times 10^{-3}$ counts s$^{-1}$.
Note that the 
3$\sigma$ upper limits during the first two years of our monitoring program from 2005
and 2006 (target ID 30334, segments 001 to 013) have been previously published in \citet{grupe07a}.

The increased number of detections especially during our intensive monitoring 
campaign in 2011 September (2011-September-01 = MJD 55805), 
as shown in the right panel of Figure\,\ref{wpvs007_xrt_uvot_lc_2010}, may suggest that
the strong absorber that is most likely the cause for the X-ray weakness of \wpvs\
 has started to disappear. It is worth noting that 
 during the intensive monitoring
campaign in 2010 June (2010-June-02 = MJD 55349)
we only got two marginal detections
(left panel of Figure\,\ref{wpvs007_xrt_uvot_lc_2010}).

In order to check the assumption that \wpvs\ has generally become brighter over
the last decade, we co-added data over long periods which are displayed as red
points in the XRT count rate light curve in Figure\,\ref{wpvs007_xrt_uvot_lc}.
These data points suggest that \wpvs\ has indeed become brighter between
2005 and about 2009, however, our recent \swift\ observations during the intense
monitoring campaign in September 2011 and thereafter indicate that this trend
did not persist, and instead, the AGN has
become fainter again. Adding all data between 2011 October and 2013 July 
(total exposure time 82 ks) results 
in a 3$\sigma$ detection at a level of 2.46$^{+2.42}_{-1.71}\times 10^{-4}$ counts s$^{-1}$.

The X-ray detections of \wpvs\ over the last years not only show strong 
flux variability, but the 
hardness ratios listed in Table\,\ref{xrt_stat} suggest strong spectral variability as well. 
While during the 'low-state' during the 2007 September observation \wpvs\ appeared to be very 
hard with a hardness ratio
HR = +0.54$^{+0.18}_{-0.12}$, during the `high state' in 2009 September 17 
the hardness ratio was 
very soft with 
HR=$-0.54^{+0.13}_{-0.16}$. Such behavior may be expected from a partial covering absorber. 
While hardness ratios can give some spectral information, 
a spectral analysis of the data is required to see if
the data are consistent with a partial covering absorber model (see below).

\subsection{X-ray spectral analysis}

As shown in Section\,\ref{x-detect}, \wpvs\ is not detected typically in a single 5ks observation. 
In order to obtain spectra, the data of several observations have to be combined. 
This method was applied to derive the photon distribution shown in \citet{grupe08b} from the \swift\ XRT 
observations in 2007 September. These data were consistent
with a neutral partial covering absorber model with an absorption column density 
of $N_{\rm H} = 1\times 10^{23}$ cm$^{-2}$ and a covering fraction $f_{\rm pc}$ = 0.95.

Table\,\ref{xrt_spec} lists the spectral analysis for detections 
when more than 30 source
counts were detected in a single observation. Due to the low number of counts in the single detections during
the intense daily monitoring campaign in 2011 September, we combined the data into one spectrum. 
Although the number of counts even during the brightest X-ray detections are low it still allows a
limited spectral analysis using Cash statistics \citep{cash79}. The analysis of these spectra 
confirms the strong spectral variability found from the hardness ratios listed in
Table\,\ref{xrt_stat}.
While the data of 2009 September 17, and 2010 July 13, are both consistent with a 
single power law model with 
a relatively flat X-ray spectral slope \ax=1.6 all other data require a partial covering absorber model. 
An F-test between the power law and partial covering absorber models of the 2007 September and 2011 
July data shows that there is a 1\%
chance that the data are drawn from a random distribution. This is, however, not the case for the 2011 
September data which clearly
require a partial covering absorber model 
with $N_{\rm H, pc} = 8.45^{+5.65}_{-2.40} \times 10^{22}$ cm$^{-2}$. Here the probability
of a random result is 0.3\%.

The spectral analysis of all these spectra was based on data when
 \wpvs\ was in relatively high
X-ray flux states. This may not represent the data when \wpvs\ is below the detection threshold. In
order to determine the spectrum in these very low X-ray flux states, we combined all
data when only 3$\sigma$ upper limits could be measured during the observations listed in
Table\,\ref{swift_xrt_3sul}. We merged all these data from 2005 October to
2013 July and obtained
a source plus background spectrum with 267 counts with a total exposure time of 434 ks. 
These data were binned with 15 counts per bin
(noted as 2005-2013 in Table\,\ref{xrt_spec}). 
Clearly, as expected, the single power law model does not fit the data. We fitted the spectrum
with a partial covering absorber model and found a high column density of $N_{\rm H}=3.6\times 10^{23}$
cm$^{-2}$ (Table\,\ref{xrt_spec}). This absorption column density is significantly higher than during
the times when \wpvs\ is detected by the \swift\ XRT, suggesting that at least in part the strong X-ray
variability that we see in this AGN is indeed caused by a change in the absorber column density.
 
A fit with the ionized partial covering absorber model {\it zxipcf} \citep{reeves08} suggests an absorber
column density of the order of $2\times 10^{23}$ cm$^{-2}$ with an ionization parameter of the order of $\xi\approx 1$
and a covering fraction of 0.96. These values (in particular, the
column density) agree roughly with the values derived from Cloudy
modeling of the absorption lines in the 2003 FUSE spectrum
 \citet{leighly09}.
Note, however, that due to the low number of counts these values are not well constrained at all.

\subsection{UVOT data}

\begin{figure}
\epsscale{0.6}
\plotone{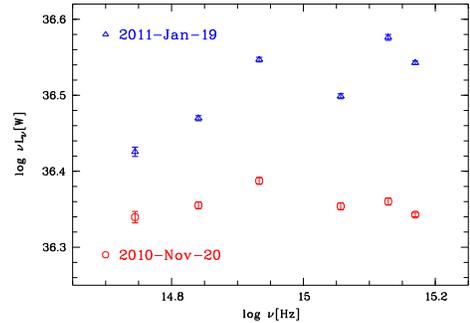}
\caption[ ] {\label{wpvs007_uv_seds} 
\swift\ UVOT Observations of \wpvs\ during the low and high state on 2010 November 20 and 2011 January 19, respectively. UVOT filter with increasing
frequency are: V, B, U, W1, M2, and W2.
}
\end{figure}

Figure\,\ref{wpvs007_xrt_uvot_lc} displays the complete UVOT light curves 
in all 6 filters since the beginning of the \swift\ monitoring
campaign in October 2005. 
The light curves during the two intensive monitoring periods in 2010 June and 2011 September 
 are displayed in Figure\,\ref{wpvs007_xrt_uvot_lc_2010}.
All magnitudes are  listed in 
 Table\,\ref{swift_results}. These data show that \wpvs\ is one of the most
  UV variable NLS1s 
 besides Mkn 335 which has shown even
 larger amplitude variations \citep{grupe12}.
Typically NLS1s  seem to vary by about 0.3 mag in the UV \citep{grupe10}.
\swift\ has monitored several AGN, especially NLS1s, over long time scales, such
as CBS 126, PG 1211+143, or PKS 0558-504 \citep[][respectively]{chaing12,
bachev09, gliozzi10, gliozzi12} but none of these has shown the amplitude
of variability as observed in Mkn 335 and \wpvs. 
 The maximum amplitude of variability seen in \wpvs\ was 0.66 mag in W2 
 within a month between 2008 June 19 and October 21
 (MJDs 54636-54760), and
 0.53 mag within two months between 2010 November 20 and 2011 January 19.
  Figure\,\ref{wpvs007_uv_seds} displays the optical/UV 
 luminosities in all six UVOT filters during these observations. 
 Clearly, not only 
 has the AGN become brighter during the 2011 January 19 
 observation, but also the UV spectral slope \auv\ has become bluer from \auv=1.00\plm0.05
  during the 2010 November 20 observation to 
 \auv=0.72\plm0.09 in the January 2011 observation. \wpvs\ varies on timescales of days. As displayed in 
Figure\,\ref{wpvs007_xrt_uvot_lc_2010} when we performed daily monitoring campaigns of \wpvs\ in June 2010 and 
September 2011,  the UV flux/magnitude changes significantly within a few days. 

Note that host galaxy contamination is an issue in low-luminosity AGN such as \wpvs. In
\citet{grupe10} we discussed a method to test the significance of the host 
galaxy contribution by
extracting the UVOT data using different source extraction radii.
 Using the standard 5$^{''}$
extraction radius and a 3$^{''}$ extraction radius (including the parameter 
apercorr=curveofgrowth in
uvotsource) results in a difference of 0.15 mag in V and 0.05 mag in W2. What this means is that we do
overestimate the brightness in V slightly, but in W2 the effect of host galaxy
 contamination is
negligible. The effect on spectral slopes such as \aox\ and \auv\ is negligible because these are
dominated by UV emission where host galaxy contribution is very low.

\begin{figure}
\epsscale{0.6}
\plotone{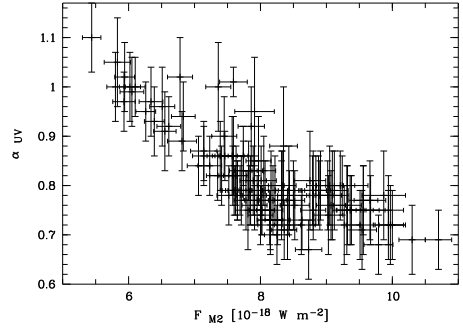}
\caption[ ] {\label{wpvs007_fm2_auv} 
Flux in the UVOT M2 filter vs optical/UV spectral slope \auv
}
\end{figure}

\begin{figure}
\epsscale{0.6}
\plotone{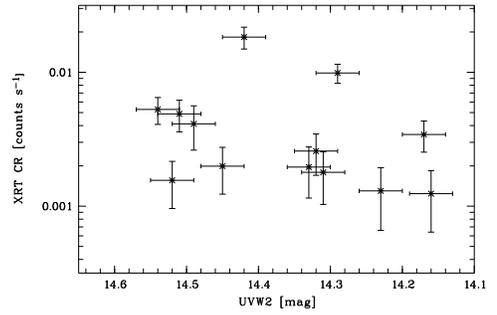}
\caption[ ] {\label{wpvs007_xrt_uvw2} 
UVOT W2 magnitude vs. XRT count rate during detections (See
Table\,\ref{xrt_stat}).
}
\end{figure}

Figure\,\ref{wpvs007_fm2_auv} shows the flux measured in the UVOT M2 filter vs. the optical/UV slope \auv. 
There is a clear anti-correlation between these two properties. The AGN becomes redder with decreasing UV flux. 
A Spearman rank order correlation analysis results in $r_{\rm s}$=--0.76 (116 data points), $T_{\rm s}$=--12.56 with a probability 
$P<10^{-8}$ of a random result. Figure\,\ref{wpvs007_xrt_uvw2} displays the
relation between the magnitude in the UVOT W2 filter\footnote{The UVOT W2 filter
was picked here because it is the bluest filter which is closer to the X-ray
emission}
 and the XRT count rate
during a detection as listed in Table\,\ref{xrt_stat}. We do not find any clear
correlation between these two properties. There is only a light trend that the
source appears to be brighter in the UV when it is fainter in X-ray. However,
this trend is not statistically significant.

 The monitoring in all 6 UVOT filters over 6 years allows to check for any time delays between different band passes.
We applied cross-correlation analysis to the data in particular the data during the daily monitoring campaigns in June 2010 and
September 2011. We did not find any significant delays between the light curves. The detections in X-rays, however, are too sparse to
allow for any cross correlation analysis between the X-ray and UVOT light curves.

\subsection{Spectral Energy Distribution}

\begin{figure}
\epsscale{0.6}
\plotone{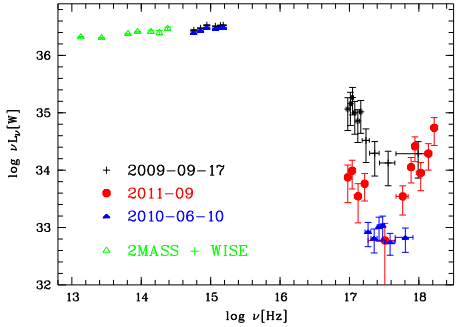}
\caption[ ] {\label{wpvs007_sed} 
Spectral Energy Distributions of \wpvs\ during the \swift\ observations from
2009 September 17 and 2011 September, and the \chandra\ observation on 2010 June 10.
Note that the X-ray data have been rebinned for display purposes, and that the
WISE and 2MASS data in the Mid and Near Infrared are non-contemporaneous.
Note that only the UVOT data from 2009 September 17 and 2010 June 10 are shown. 
}
\end{figure}

With several X-ray detections of \wpvs\ we are able to study the
variability of the spectral energy distribution (SED). Figure\,\ref{wpvs007_sed} displays the
SEDs during the 2009 September 17 `high state' and the 'low state' observations during the 
2010 June 10 \chandra\ observation\footnote{The full analysis of the \chandra\
data will be presented in Cooper et al. (in prep.)} 
and the
2011 September (combined data)
\swift\ detection. In addition this plot also made use of near infrared data from the 
Two Micron All Sky Survey (2MASS) and mid-infrared data from the Wide-field Infrared Survey Explorer (WISE).
The fluxes and luminosities of the 2MASS and WISE measurements are listed in Table\,\ref{wpvs_ir}.

All optical to X-ray spectral slopes \aox\ for each 
X-ray detection of \wpvs\ are listed in Table\,\ref{xrt_stat}\footnote{Except for the 2007 September 
observation when the \swift\ UVOT was still turned off during the gyro
 recovery phase \citep[See][for details]{grupe08b}}. 
During the \chandra\ observation on 2010 June 10 and the \swift\ detection in 2011 September
the optical to X-ray spectral
slope \aox\footnote{\aox=--0.384 $\log(f_{\rm 2keV}/f_{2500\AA}$) as defined by
\citet{tananbaum79}. The uncertainties in \aox are 0.2} was 2.49.
 Following the definition by \citet{brandt00}, at those times
\wpvs\ was an X-ray weak AGN. During the 2009 September 17 `high state', however,  the slope
was \aox=1.89. This value is still significantly larger than 
expected from a low-luminosity AGN such as \wpvs. According to the \aox - log
$l_{2500 \AA}$ relation given in \citet{grupe10} (equation 12) we would have
expected the optical to X-ray spectral slope to be about \aox=1.26.
This is the same value that can be derived from the relation given by
\citet{just07} and \citet{strateva05}. 
 Even when corrected for absorption, \wpvs\ still remains 
rather X-ray quiet considering its UV luminosity density.
The changes in the optical to X-ray spectral slopes \aox\ become apparent when looking at
Figure\,\ref{wpvs007_sed}. These changes are primarily driven by the strong changes in the 
X-ray spectra. 
Note that there is no correlation between the optical-to-X-ray spectral slope \aox\ and the
UV spectral slope \auv. In general, there is a strong anti-correlation between \aox\ and \auv\ found
among AGN as shown from the \swift\ observations by \citet{grupe10}. However, all optical-to-X-ray
spectral slopes measured for \wpvs\ are outside the \aox\ range shown in \citet{grupe10}. From its
\auv\ of the order of +0.80 one would expect the \aox\ to be about 1.3, which is obviously not the
case. 

In \citet{grupe10} we fitted optical to X-ray part of the
SEDs with a power law with exponential cutoff model to represent
the UV/EUV part of the spectrum and an absorbed power law model in the X-ray regime. 
However, although this model can describe the data, it is not a physical model and in some cases 
leads to overestimating the bolometric luminosity. Recently \citet{done12} introduced a spectral
fit model {\it optxagnf} which models the spectrum of a thin accretion disk including a 
color correction of the blackbody component, comptonization in the disk to obtain the soft
X-ray excess, and inverse Compton up-scattering to form the hard X-ray component. This model
is included in the most recent version of {\it XSPEC} 12.7.1. 

We applied this model to the 2009 September 17 `high state' data. 
Due to the low number of X-ray counts in that observation we can use this model
only for a consistency check.
We measured a bolometric luminosity 
of log $L_{\rm bol} = 6\times 10^{37}$ W (equivalent to $6\times 10^{44}$ erg
s$^{-1}$).
With its black hole mass of $4\times 10^6$ \msun\ 
\citep{leighly09}
this means that 
\wpvs\ is accreting at the Eddington limit. Note, that this is a factor of 10
higher than what was estimated by \citet{leighly09} based on the 1996 HST FOS
spectrum. However, this is consistent with the long-term X-ray light curve shown
in Figure\,\ref{wpvs007_xray_lc}. 

\subsection{Near Infrared Spectroscopy}

\begin{figure*}
\epsscale{1.6}
\plottwo{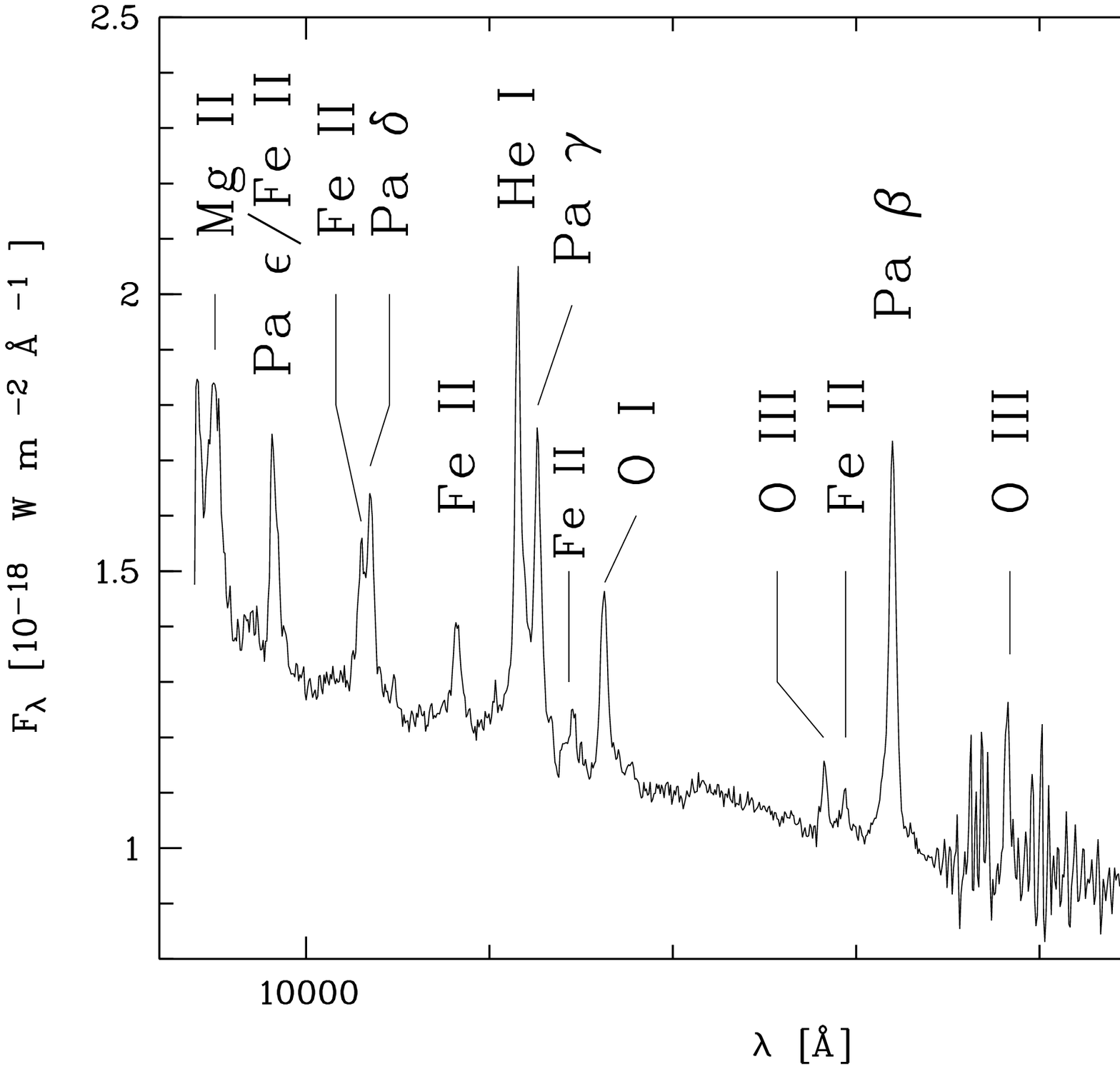}{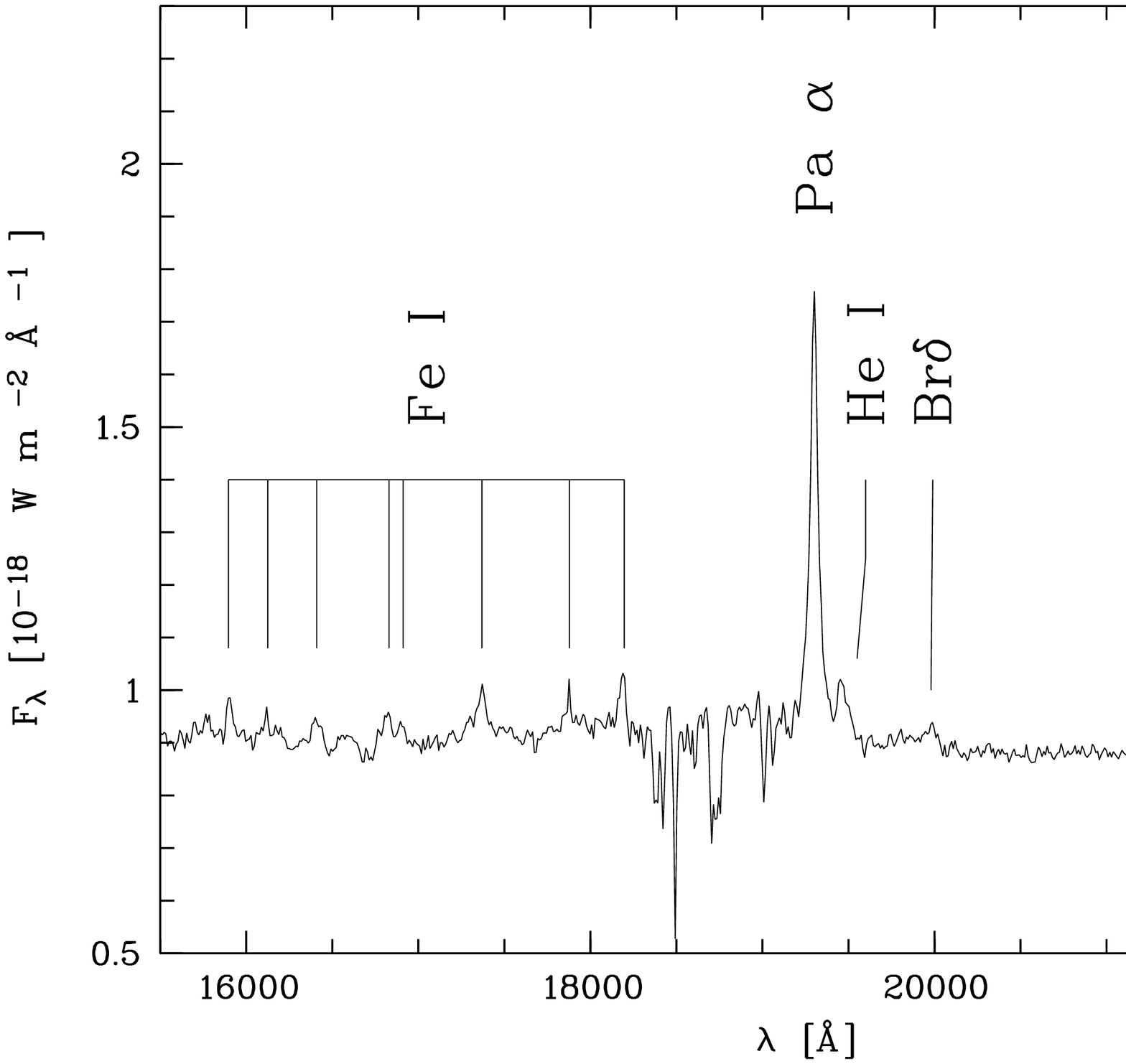}
\caption[ ] {\label{wpvs007_nir_spec} 
Near Infrared spectrum of \wpvs\ in the J plus H and H plus K wavelength bands observed on 
2004 September 12 with the 3.5m ESO NTT. Note that the noise around 1.4$\mu$m and short of Pa$\alpha$ 
is due to correction for atmospheric absorption.
}
\end{figure*}

Figure\,\ref{wpvs007_nir_spec} displays the 2004 NTT SOFI NIR spectrum of \wpvs\ 
in the J plus H and H plus K  bands. It is rich in emission lines. 
Transitions were identified following the NIR spectra of 
NLS1s published by \citet{landt08, garcia12} and the line
catalogue of the National Institute of Standards \citep[NIST, ][]{ralchenko11}.
As pointed
out by \citet{landt08}, NIR spectroscopy of Seyfert 1 galaxies is a relatively new field and only some brighter AGN
have been studied in the NIR \citep[e.g.,][]{rodriguez11, grupe02}. \wpvs\ exhibits very strong FeII and FeI emission
lines, even stronger than in the NLS1 prototype I Zw 1 \citep[see ][]{garcia12}. 
All identified emission lines in the NIR spectra are 
listed in Table\,\ref{nir_lines} with their Full Width at Half Maximum (FWHM) 
and line flux. 
The equivalent widths of the Pa$\alpha$ and
Br$\gamma$ lines are 68 and 7.6\AA, respectively. Compared with other NLS1s, as,  
e.g., listed in \citet{grupe02}, these values
appear to be at the higher end. 

%% Discussed later; mentioned out of context here; therefore removed. 
%% The near-IR spectrum was obtained on 2004 September 12.  Less than one
%% year before, on 2003 November 6, we observed WPVS 007 using FUSE 
%% \citep{leighly09}.
%% The FUSE spectrum revealed that WPVS 007 had
%% developed a BAL outflow with $V_{max}\sim 6,000\, \rm km\, s^{-1}$
%% since the 1996 HST FOS observation.  The FUSE spectrum contained
%% \ion{P}{5} absorption lines.  Due to the low abundance of phosphorus,
%% \ion{P}{5} is observed only when the column density of the outflow is
%% relatively high (Hamann 1998), as well as being relatively highly
%% ionized (Leighly et al. 2009).

\subsection{Optical spectroscopy}
\begin{figure*}
\epsscale{1.6}
\plottwo{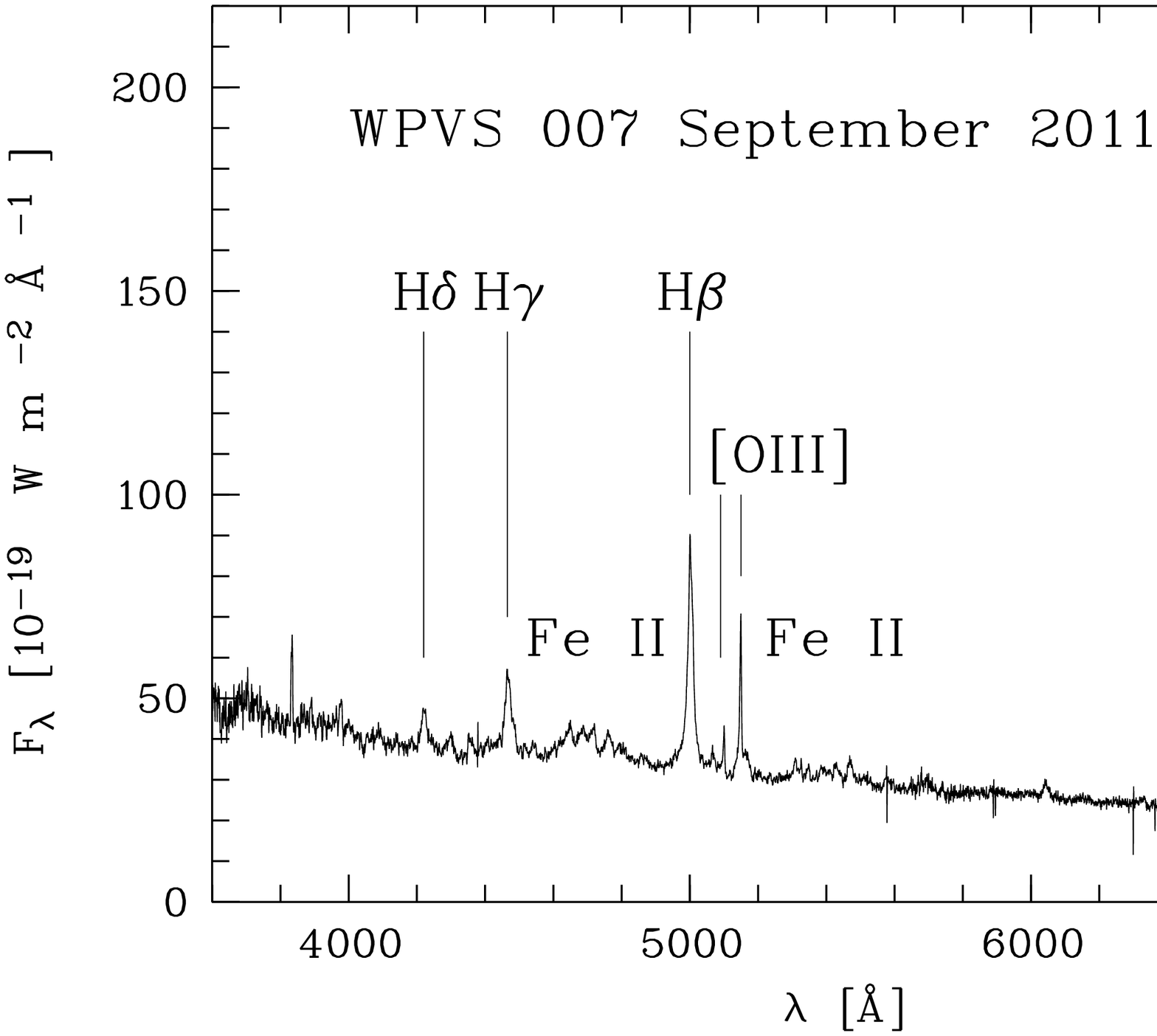}{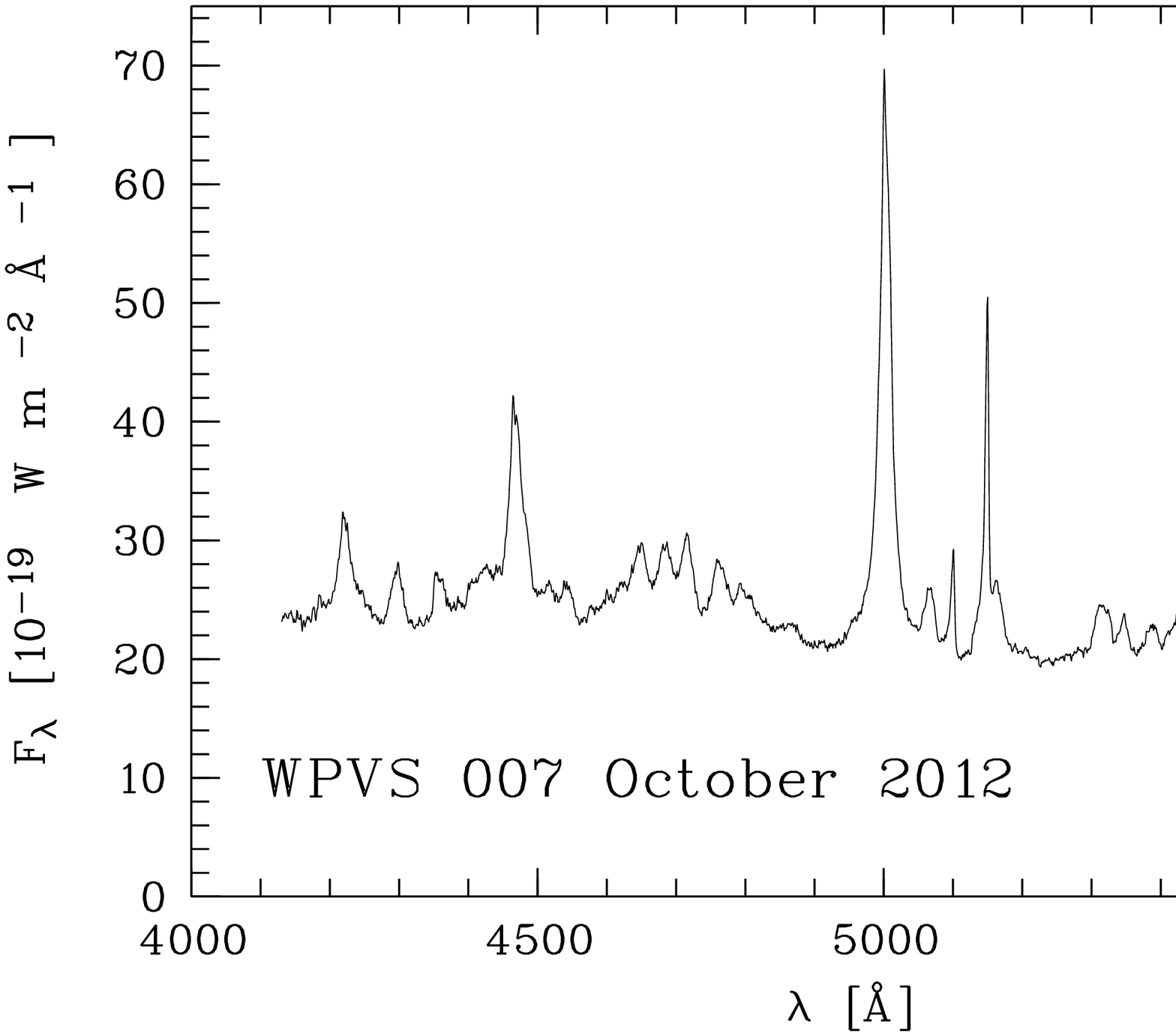}
\caption[ ] {\label{wpvs007_opt_spec} 
Optical spectrum of WPVS 007 obtained in September 2011 at the ANU 2.3m 
telescope at Siding Spring Observatory using the 3$^{"}$ extraction radius,
 and the 4.1m SOAR telescope  in October 2012
(left and right panels, respectively).
}
\end{figure*}

Figure\,\ref{wpvs007_opt_spec} displays the most recent optical 
spectra obtained for \wpvs\ at the 
2.3m ANU telescope in September 2011,
and the 4.1m SOAR telescope on Cerro Pachon in 
Chile in October 2012 (left and right panels, respectively).
The reason for taking these optical spectra was to see, whether
any changes in the emission lines had occurred (since the 1990s), given the strong
apparent changes of observed (line-of-sight) continuum emission
(note that FeII variability has been reported previously by, e.g., 
\citet{giannuzzo96} and \citet{shapovalova12}). 
 
For the 2011 September spectrum, which was taken with the Integrated Field 
Spectrograph, we extracted a nuclear spectrum with 1" extraction radius. 
%% this minimizes the host contribution. 
We do not find significant changes in the FeII fluxes 
over more than a decade. 
Spectra taken in 2007 at
CTIO \citep{dunn08}, and most recently with the 4.1m SOAR telescope, 
confirm that there are no significant changes in the emission line strengths
compared with the spectra taken in 1992 and 1995 (the properties of the 
optical emission lines measured at that time were given in \citet{grupe99}). 
This result then implies that no dramatic changes have occurred in the
 photoionizing continuum seen by the Balmer-line emitting gas

\section{\label{discuss} Discussion}

\subsection{On the nature of the X-ray low-state of WPVS 007} 

We presented new \swift\ observations on the X-ray transient NLS1 \wpvs\ 
covering a period between October 2005 \citep{grupe07a,grupe08b} and 
March 2013, in order to see whether 
the X-ray faintness of WPVS 007 persisted, or whether, and in which
way, the source re-brightened again. 
We find, that WPVS 007 remains below the \swift\ detection limit 
most of the time, but shows some periods of short-term fluctuations in
X-ray flux, increasing almost to the initial ROSAT high-state 
once in 2009.  

A key question then concerns the nature of the X-ray faintness
of WPVS 007:  is its X-ray weakness
intrinsic,  or caused by absorption along the line of sight
and related to the strong UV BAL flow which developed in the UV ? 
What causes the short-timescale high-amplitude fluctuations we
see with \swift\ ?  

We begin with a short summary of X-ray weak AGN and models suggested
to explain them, and then move on to exploring the link
between the long-lasting X-ray low-state of WPVS 007 and its
strong UV BALs.  

While NLS1s as a population typically appear to be bright in X-rays, 
their high \lledd\ 
and steep X-ray spectral slope \ax\ result in a steeper optical-to-X-ray 
spectral slope \aox, which 
makes them appear to be 
X-ray weaker when compared with Broad Line Seyfert 1s \citep[e.g.,][]{grupe10} 
when comparing the X-ray to
the UV flux.
However, generally speaking they do
not appear to be X-ray weak according to the definition of X-ray weakness 
of \aox$>$2.0 as given by \citet{brandt00}. 
Nevertheless X-ray weak NLS1s are known \citep[e.g.,][]{williams02, williams04}.  
Among  the best-studied examples are  
PHL 1811 \citep{leighly07} and PHL 1092 \citep{gallo04b, miniutti12}. 
X-ray weak quasars have even been reported at high 
redshifts \citep{wu11}. 
Therefore, X-ray weakness may not be so uncommon even 
among NLS1s. 

Recently \citet{miniutti12} 
reported a drop in the soft X-ray flux in PHL 1092 by a factor of 260 between 2003 and 2008. 
This is a drop in X-ray flux similar
to \wpvs. PHL 1811 has an \aox=2.3 \citep{leighly07} which is  similar to  
\wpvs\ when detected. 
PHL 1092 displays variability in the X-ray loudness
between \aox=1.6 and 2.5 \citep{miniutti12},
which again is similar to the variability in \aox\ we have 
found in \wpvs\ (see Table\,\ref{xrt_stat}). 
 X-ray variability in other AGN, and NLS1 galaxies in particular, of much lower
amplitude has typically been interpreted in terms of either absorption or
reflection, or both. Example seen by \xmm\ as well as Suzaku are:
PG 2112+059 \citep{schartel10}, PG 0844+349 \citep{gallo11}, MCG-6-30-15 \citet{inoue11},
NGC 4051 \citep{lobban11}, Mkn 335 \citep{grupe12}, Fairall 9 \citet{lohfink12},
1H0707--495 \citep{miller10, fabian12, dauser12}, and RX J2340--5329 \citep{schartel13}.

However, \wpvs\ is different from all these X-ray weak AGN and absorbed AGN.  
As discussed by \citet{leighly07} for PHL 1811, 
the X-ray spectrum of this NLS1 can be fitted by a single power 
law with just Galactic absorption with no evidence for
an intrinsic absorber in the line of sight. This, however, is not the case in  
\wpvs. Its X-ray spectra can simply not be fit by a single power law model.

We also note that   
none of the X-ray weak NLS1 galaxies, including PHL 1092 and PHL 1811,
do  show any signs of UV absorption lines \citep{leighly07, leighly07b}.
As we know, this is certainly not the case in \wpvs\,
where strong, deep BAL troughs have been found \citep{leighly09}. 
The detection of these strong BAL troughs in the UV therefore
strongly suggests that the deep and long-lasting X-ray low-state of WPVS 007 
is caused by absorption. Fluctuations in the X-ray count rate
could then be caused by clumpiness of this absorber as it 
crosses the line of sight{\footnote{ 
Note that although variable BALs are common, the transitional change as observed
in \wpvs\ is extremely rare 
 \citep[e.g.,][]{hamann08, capellupo11, filiz12}.}}

We have shown that the fits to the X-ray data of \wpvs\ significantly 
improve when a partial covering absorber model is applied instead of a 
single power law model (see Table\,\ref{xrt_spec}) 
{\footnote{In general, the curvature found in 
 X-ray spectra, such as seen here in \wpvs\ ,
 can  also be modeled by a blurred reflection
model; as we have demonstrated for the X-ray faint and intermediate states in the 
NLS1 Mkn 335 \citep{grupe07b, grupe08, grupe12, gallo12}. Mkn 335,
however, lacks strong UV absorption features. }}. 
A large absorption column
density of about $5\times 10^{23}$ cm$^{-2}$ is inferred, 
when the AGN is not detectable in single \swift\ XRT observations. 
 During times when \wpvs\ is detected, the 
absorption column density is significantly lower (see Table\,\ref{xrt_spec}).
To summarize, the X-ray variability observed in \wpvs\ is consistent 
with a variable partial covering absorber in the 
line of sight.

The column density of the absorber, inferred from UV observations,
is on the order of $2\times 10^{23}$ cm$^{-2}$, with a 
photoionization parameter $\xi$ of the order of 1, assuming 
solar metallicity.
Since the UV absorber
is ionized, and in order to compare more directly with the X-rays, we have
re-fitted the Swift data derived from the observations when \wpvs\ was not detected
with the {\em ionized} partial covering absorber model {\it zxipcf} \citep{reeves08}.
We find that the absorption column density and ionization parameter derived from this spectrum agrees with the 
2003 FUSE data. Deep {\em {simultaneous}}
UV and X-ray spectroscopy
observations would be required, however, in order to perform a more rigorous
comparison.

\subsection{UV variability}

\begin{figure*}
\epsscale{2.2}
\plottwo{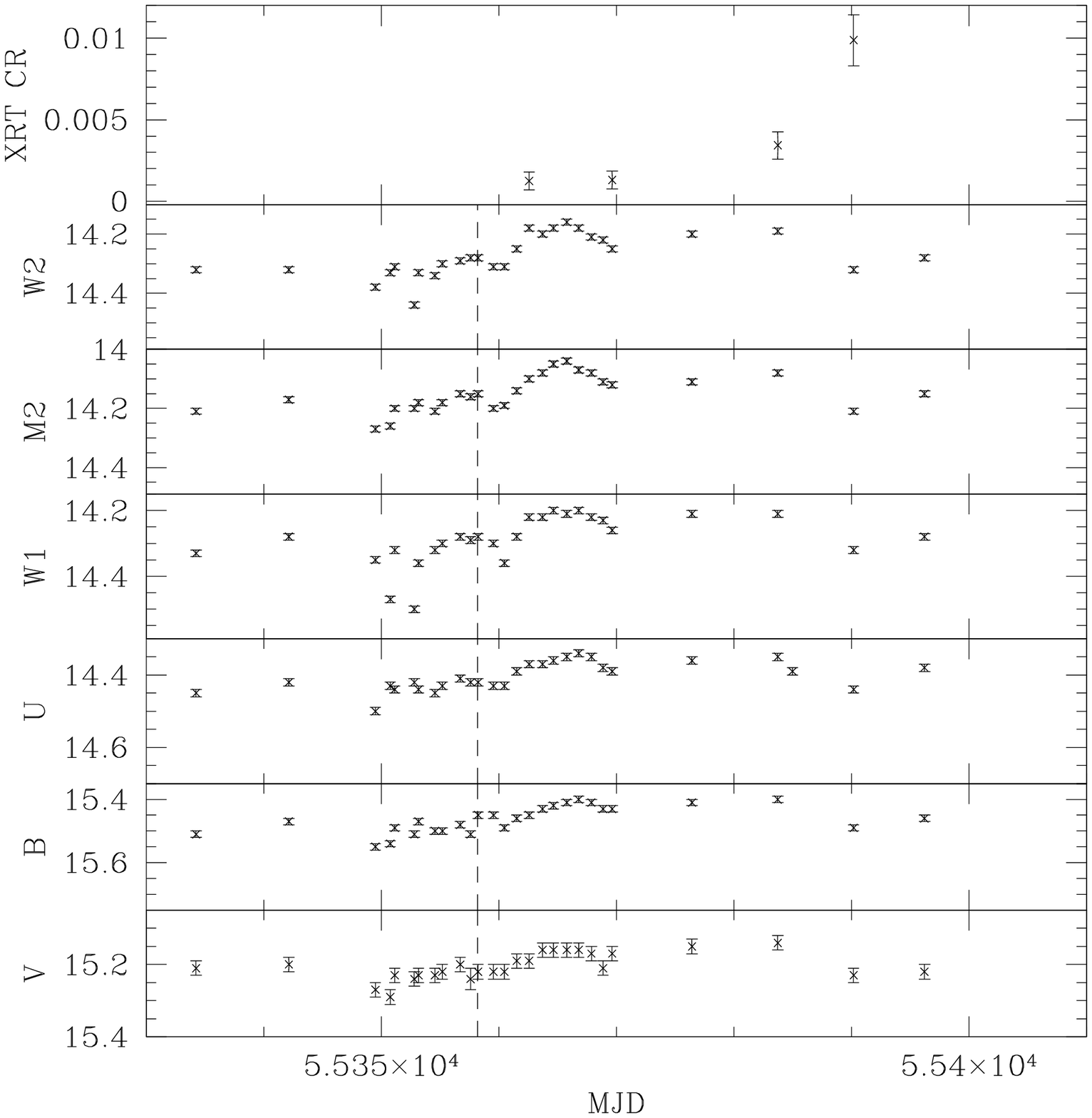}{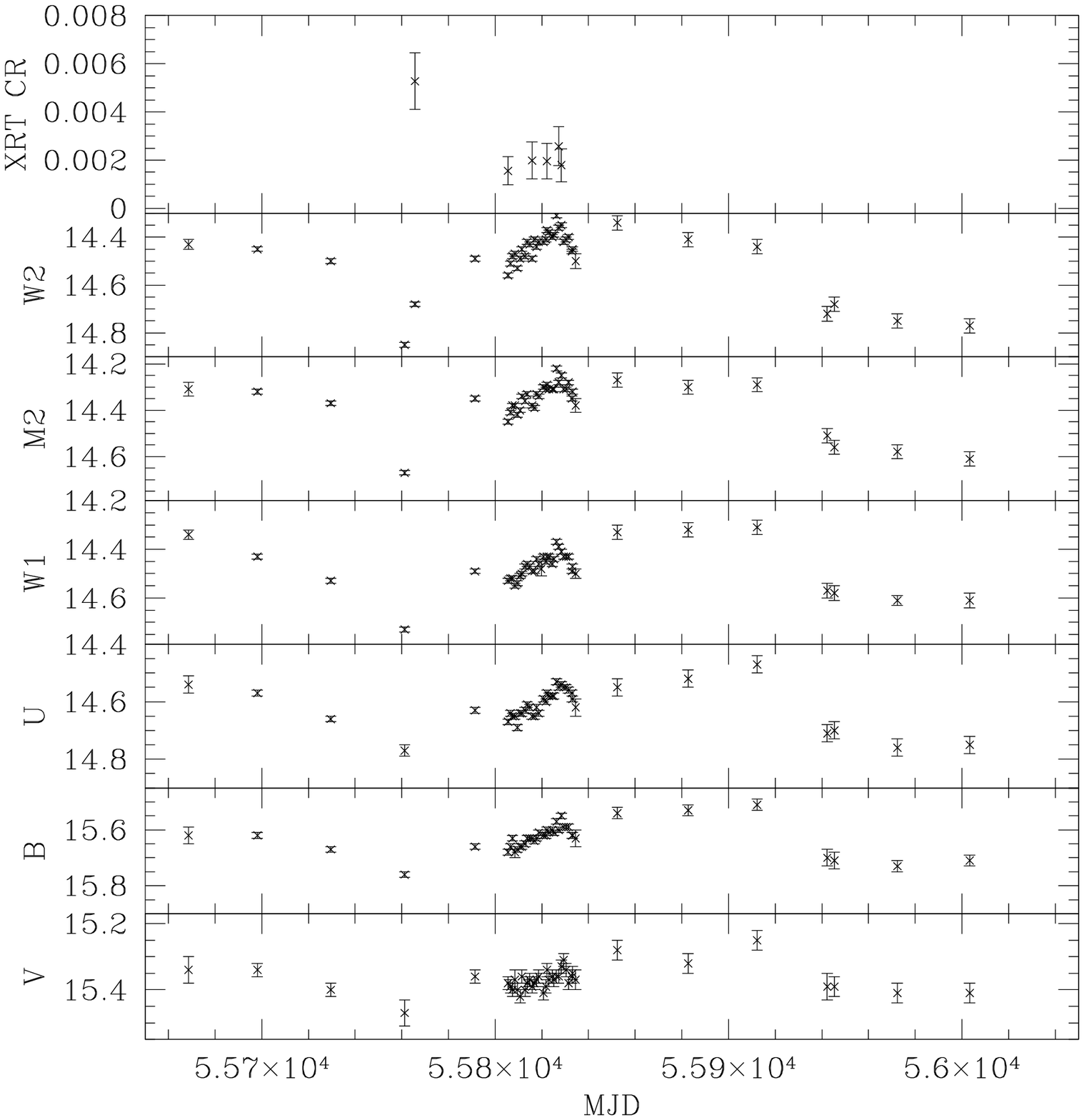}
\caption{\label{wpvs007_xrt_uvot_lc_2010} XRT and UVOT light curves during and
around the intense monitoring campaigns in June 2010 and September 2011 (2010-June-02 = MJD 55349, and 
2011-September-01 = MJD 55805).
Note
that only detections in the XRT are shown. During all other observations only
3$\sigma$ upper limits can be given as shown in Figure\,\ref{wpvs007_xray_ul}.
}
\end{figure*}

As shown in the \swift\ UVOT light curve (Figures\,\ref{wpvs007_xrt_uvot_lc} and
\ref{wpvs007_xrt_uvot_lc_2010}), \wpvs\ varies strongly in the UV on long as
well as intermediate time scales. 

 As for the optical/UV we noticed  a strong anti-correlation between 
 the flux measured in the UV (M2 filter) and the
 optical/UV spectral slope \auv. There are several ways to 
 explain such an anti-correlation. For instance, (a)  intrinsically by a change 
 in the accretion rate and therefore \lledd, or (b) by a change in the 
 reddening. As shown by \citet{grupe10}, there is a clear correlation 
 between the Eddington ratio \lledd\ and the UV spectral slope \auv. 
 If we estimate \lledd\ based on the extreme values of \auv\ of about 
 1.1 and 0.7 we find \lledd\ ratios of 0.54 and 0.95, respectively. 
 Alternatively, to cause a change in the optical/UV continuum as shown in 
 Figure\,\ref{wpvs007_uv_seds} only requires a change in the 
 intrinsic $E_{\rm B-V}$ by 0.06 mag. 
 We note in passing that the UV absorption lines are outside the wavelength
 windows of the UVOT filters, and so we are insensitive to any changes 
 in the UV absorption lines. 

Dusty clouds in motion, passing our line-of-sight, may therefore 
be responsible for the observed UV variability.
 Dust does not survive temperatures higher
than 1500K. The dust sublimation radius (of a dusty medium around 
the central black hole) 
 can be estimated by the relation
given in, e.g., \citet{kishimoto11} with $R_{\rm sub} = 0.47 \times ( 6 \times
\nu L_{\rm 5500\AA})^{0.5}$ where
 $L_{\rm 5500\AA}$ is the luminosity at 5500\AA\ in
units of $10^{39}$ W and  $R_{\rm sub}$ the dust sublimation 
radius in units of pc. 
Using the optical spectra of \wpvs\ we can estimate the inner radius of the
torus of the order of about 20 light days. 

If we assume that \wpvs\ is similar in structure to a BAL QSO,  then the wind of
\wpvs\ has two components, a gaseous, dust free component that is responsible
for the absorption in X-rays and the UV broad absorption lines, and a dusty
component which is at larger radii with $R > R_{\rm sub} ~ 20$ ld (0.05 pc). 
%% This is similar, for example to  
%% what has been suggest for protostars as shown in the model by
%% \citet{bans12}. 
The  broad line region
gas on the other hand in \wpvs\
 is located at $R_{\rm BLR}$ = 5 ld following the
relation by \citet{kaspi00}. This is significantly smaller than the dust
sublimation radius. (This, then, of course, can explain why we do see 
significant variability in the UV continuum, caused by dust, 
without affecting the photoionizing
continuum that is seen by the BLR.)  

At a distance of 20 ld, the velocity of the dust cloud around the central black
hole will be of the order of 1000 km s$^{-1}$.  The UVOT light curves
shown in Figure\,\ref{wpvs007_xrt_uvot_lc_2010}, in particular the light curves
during the intensive daily monitoring campaign in September 2011, show that
\wpvs\ can vary significantly within a few days. For a high \lledd\ AGN such as 
\wpvs\ we can reasonably  assume a standard $\alpha$ accretion disk.
The temperature stratification of an $\alpha$
disk is described  in \citet{frank02} and in a more appropriate form for an AGN in
\citet{peterson97}, eq. 3.20. 

 The central wavelength of the W2 filter in the UVOT is at
  1923\AA\ \citet{breeveld10} which is equivalent to a
temperature of 75000K.  If we assume that \wpvs\ accretes at \lledd=1 and that the
mass of the central black hole is $4\times 10^{6}$\msun \citep{leighly09}, 
then the UV emission in
the W2 filter is emitted at $r_{\rm UV} = 50 R_{\rm S}$ or $5.8\times 10^{11}$m. 
%% At the distance of the absorber, this
%% diameter is equivalent to 8$^{'}$ in the sky. With a velocity of 1000 km
%% s$^{-1}$, the dust clouds needs $3.26\times 10^9$ s or roughly 100 years for one
%% orbit. This also means that the angular velocity of the dust cloud is 4$\times
%% 10^{-4}$ $^{''}$ s$^{-1}$.
Therefore, a dusty cloud needs roughly two weeks to cover the
UV emitting region.  This is
consistent  within the uncertainties with the variability time scale we see from
the UVOT light curves (Figure\,\ref{wpvs007_xrt_uvot_lc_2010}), which suggest a 
time scale of about a week.

\subsection{NIR spectroscopy}

\subsubsection{Emission lines} 

We have reported for the first time on NIR spectroscopy of WPVS 007. 
The NIR spectra display a variety of strong NIR emission lines as 
shown in Figure\,\ref{wpvs007_nir_spec} and listed in 
Table\,\ref{nir_lines}. We compared our NIR spectra of WPVS 007 with 
NIR spectra of the AGN sample of \citet{landt08}, and, in particular with the galaxies  
Mkn 335, Mkn 110, and PG 0844+349. This comparison shows that the NIR spectra of 
WPVS 007 are quite normal for a NLS1 galaxy. 
Although the Fe emission in WPVS 007 appears to be stronger than in 
Mkn 335 and PG 0844+349, it is not as strong as seen in Mkn 110
\citep[see spectra shown in][]{landt11}.  

\subsubsection{Absorption lines}
 
As discussed in \citet{leighly11} \citep[also, e.g., Fig.\ 1 in][]{leighly12}, 
the observation of \ion{P}{5} predicts the presence of
absorption by \ion{He}{1}*$\lambda 10833$. Although this object is
known to have particularly variable BALs, BAL variability is
typically modest (e.g., Capellupo et al.\ 2012), so the IR and FUSE
observations (September 2004 and November 2003, respectively)
can be treated as roughly contemporaneous. In addition,
the IR observation was performed after the FUSE observation.
Therefore, it is probably reasonable to suppose that the outflow was
still present for the IR observation, with roughly the same opacity
profile. However, we see no evidence of the predicted \ion{He}{1}
absorption line in the IR spectrum. 

We used the optical depth profile derived from the FUSE \ion{P}{5}
line (Leighly et al.\ 2009, Fig\ 4) to obtain an upper limit on the
\ion{He}{1}* apparent column density, using IRAF {\tt Specfit}.  The
spectrum in the region of the \ion{He}{1}*$\lambda 10833$ emission
line was modeled using a power law continuum and several Lorentzian
line profiles.  Specifically, in the vicinity of the putative
absorption, our model included the following emission lines:
\ion{Fe}{2}$\lambda\lambda 9956,9998$, \ion{H}{1}$\lambda 10052$,
\ion{Fe}{2}$\lambda\lambda 10491,10502$, \ion{He}{1}*$\lambda 10833$,
\ion{H}{1}$\lambda 10941$, \ion{Fe}{2}$\lambda 11126$, and
\ion{O}{1}$\lambda 11290$.  We also included a tiny line located at
$1.0713\rm \, \mu m$.  Before including absorption, the width of this
line was $1150 \pm 430 \rm km\, s^{-1}$, comparable to the other
lines, and the equivalent width was $\sim 3.5$\,\AA\/.  This line
remains unidentified; it is quite near an unidentified line in Ark~564
near 1.0740 \AA\/ (Landt et al.\ 2008), but it is sufficiently far
away to be a different line.  The IR spectra of most of the other
nearby Seyfert 1 galaxies published by Landt et al.\ 2008 generally
show no emission lines between \ion{Fe}{2}$\lambda\lambda 10491,10502$
and ion{He}{1}*$\lambda 10833$.  We included this unidentified line in
our model, fixing the width and centroid, but leaving the
normalization free.

Adding the absorption trough to the model, we estimated the upper
limit in the \ion{He}{1}* column density in two ways.  First, we
increased the normalization of the absorption line until the $\chi^2$
increases by 6.63 after refitting, appropriate for 99\% confidence for
one parameter of interest.  This yielded a log \ion{He}{1}* column
density of 12.7 for both the case when the continuum only is absorbed,
and when both the emission lines and continuum are absorbed.

We also evaluated the upper limit more conservatively by increasing
the depth of the absorption until the difference between the refit
model and the data was comparable to the error bars on the spectrum.
This criterion can only be sensibly satisfied between the \ion{He}{1}*
emission line and the unidentified emission line, constraining our
examination to the region of the absorption between $-2760$ to
$-1480\rm \, km\, s^{-1}$.  As shown in Fig.\ 4 of Leighly et al.\
2009, the optical depth is highest in this velocity range of the
outflow. This procedure yielded a conservative upper limit on the log
\ion{He}{1}* column density of 12.9, with $\Delta \chi^2$ of 22.8 and
19.5 for absorbed continuum, and absorbed continuum plus lines,
respectively. 
Both of these estimates are much lower than the log \ion{P}{5} column
of $\sim 15.4$ obtained by \citet{leighly09}.  The implications of
this dramatic difference will be discussed by Cooper et al.\ (in prep.).

In the future, additional NIR spectra of WPVS 007 will allow us to see
whether any NIR features (emission or absorption lines) varied.

\subsection{Optical spectroscopic monitoring}

Although \wpvs\ has exhibited extreme variability in X-rays as well as in the
UV/optical continuum, we have shown that its optical emission lines
did not exhibit any
significant changes over the last two decades. 
This result implies, that the
photoionizing continuum seen by the emission-line gas has not changed
significantly over at least the last 20 years. 
We can therefore assume that the
intrinsic spectral energy distribution of \wpvs\ has not changed much. 
This
finding confirms the absorption scenario that we infer from our UV and X-ray
observations of \wpvs, and so most of the absorption just occurs along our line-of-sight.  

\section{Conclusions}

We presented the results from our long-term monitoring campaign of \wpvs\ between 
 October 2005 and
July 2013. This has been the
longest continuous \swift\ monitoring campaign of any AGN. Our main findings are
as follows: 

\begin{itemize}
\item While remaining X-ray faint most of the time, WPVS 007 shows some
remarkable fluctuations in X-ray flux during the \swift\ monitoring
campaign between October 2005 and April 2013. During this period 
it has shown variability by factors of at least 30.
In particular, it temporarily almost reached its initial
ROSAT high state during one epoch in September 2009.
 
\item The X-ray weakness seen in \wpvs\ can be explained by X-ray absorption,
due to partial covering with an absorber of varying column density.   
During the times
when the AGN is not detectable in individual observations by the \swift\ XRT,
this absorption is of the order of 
$4\times 10^{23}$ cm$^{-2}$. 
The lower absorption column densities measured during the individual X-ray detections 
suggest that the strong
X-ray variability is primarily caused by an absorber, most-likely an 
outflow in the line of sight which
is consistent with the BALs found in the UV spectra of \wpvs.
Clumpiness in this flow might be the reason for the large-amplitude
X-ray variability seen with \swift\ . 

\item Applying an ionized partial covering absorber model to the X-ray data merged from all
non-detections, implies that the column density and ionization
parameter of the 
X-ray absorber is comparable with the absorber found from the UV FUSE spectra,
suggesting a common origin.  

\item WPVS 007 is one of the most variable AGN in the UV. This is especially  
unusual for a NLS1. The highest amplitude of UV variability amounts
to 0.66 mag within two months. 

\item The UV spectral variability might be caused by extinction
by dusty clouds or a dusty wind.
The scenario is consistent with clouds beyond the dust sublimation
 radius in \wpvs\, which is of the order of 20 ld (0.05 pc).

\item We presented NIR spectra for the first time of \wpvs. The rich
emission-line spectra 
show strong Fe emission lines, neutral and singly
ionized. The NIR data also show  MgII, O I, OIII, and He II emission lines.
Emission-line ratios are not unusual for this class of AGN. 
Analyzing the NIR spectra showed that the absorption column densities derived from these data
are significantly lower than what has been found from the UV and X-ray data.

\item New optical spectra obtained in 2011 and 2012 show that the strengths of the 
optical emission lines have not changed in comparison with optical data taken in the mid 1990s,
implying that  
the underlying photoionizing continuum seen by the emission-line clouds has not changed over 
the last two decades.

\item \wpvs\ represents an important link between BAL QSOs and NLS1s. 
In particular, it is a
low-luminosity, low-redshift AGN that exhibits BALs 
that are typically present only in high-luminosity quasars at higher redshifts. 
Timescales in \wpvs\ are therefore shorter, 
making this nearby AGN an important laboratory for understanding outflows in AGN, and in 
BAL QSOs in particular. 
\end{itemize}

\acknowledgments

We would like to thank Neil Gehrels for approving our ToO requests and
the \swift\ team for performing the ToO observations of
WPVS 007 and scheduling the AGN on a regular basis. 
Secondly, we wish to thank Kim Page for carefully reading the manuscript and for useful comments and suggestions.
We would also like to thank the anonymous referee for useful comments and suggestions. 
Many thanks to Chris Done for her help with any questions we had regarding 
her {\it optxagnf} SED model. We want to especially thank Hermine Landt for providing the
NIR spectra of her AGN sample \citep{landt08, landt11} and Michael Crenshaw for
making his 2007 CTIO optical spectrum \citep{dunn08} available to us.
This research was supported by the DFG cluster of 
excellence ``Origin and Structure of the
Universe'' (www.universe$-$cluster.de). 
SK would like to thank the Aspen Center for Physics for their
hospitality. The Aspen Center for Physics is supported by NSF
grant 1066293.
This research has made use of the NASA/IPAC Extragalactic
Database (NED) which is operated by the Jet Propulsion Laboratory,
Caltech, under contract with the National Aeronautics and Space
Administration. This publication makes use of data products from the 
Wide-field Infrared Survey Explorer, which is a joint project of the University of 
California, Los Angeles, and the Jet Propulsion Laboratory/California Institute of Technology, 
funded by the National Aeronautics and Space Administration. 
This publication makes use of data products from the Two Micron All Sky Survey, which is
 a joint project of the University of Massachusetts and the Infrared Processing and Analysis
  Center/California Institute of Technology, funded by the National 
Aeronautics and Space Administration 
  and the National Science Foundation. Based on observations obtained at 
  the Southern Astrophysical Research (SOAR) telescope, which is a joint project 
  of the Minist\'{e}rio da Ci\^{e}ncia, Tecnologia, e Inova\c{c}\~{a}o (MCTI) 
  da Rep\'{u}blica Federativa do Brasil, the U.S. National Optical Astronomy 
  Observatory (NOAO), the University of North Carolina at Chapel Hill (UNC), 
  and Michigan State University (MSU).
This research has made use of the
  XRT Data Analysis Software (XRTDAS) developed under the responsibility
  of the ASI Science Data Center (ASDC), Italy.
\swift\ at PSU is supported by NASA contract NAS5-00136.
This research was also supported by NASA contracts NNX07AH67G,  NNX10AK85G, 
NNX10AF49G, and NNX11AF82G.
(D.G.).

\clearpage

%Figure 1

% Figure 2

% Figure 3

% Figure 4

%Figure 5

%Figure 6

%Figure 8

% Figure 09

%Figure 7 -- new figure 9

% Figure 010

\clearpage

\begin{deluxetable}{cccccrrrrrrr}
\tabletypesize{\tiny}
\tablecaption{\swift~ Observation log of WPVS 007 after January 2008\tablenotemark{1}
\label{swift_log}}
\tablewidth{0pt}
\tablehead{
\colhead{ObsID} &
\colhead{Segment} &
 \colhead{T-start\tablenotemark{2}} & 
\colhead{T-stop\tablenotemark{2}} 
& \colhead{MJD} 
& \colhead{$\rm T_{\rm XRT}$\tablenotemark{3}} 
& \colhead{$\rm T_{\rm V}$\tablenotemark{3}} 
& \colhead{$\rm T_{\rm B}$\tablenotemark{3}} 
& \colhead{$\rm T_{\rm U}$\tablenotemark{3}} 
& \colhead{$\rm T_{\rm UVW1}$\tablenotemark{3}} 
& \colhead{$\rm T_{\rm UVM2}$\tablenotemark{3}} 
& \colhead{$\rm T_{\rm UVW2}$\tablenotemark{3}} 
} 
\startdata
30334 & 030 & 2008-01-12 17:56 & 2008-01-12 22:55 & 54477.85 &   1528 &  144 &  156 &  156 &  310 & 260 & 623 \\ 
      & 031 & 2008-06-19 00:03 & 2008-06-19 06:42 & 54636.14 &   3753 &  334 &  334 &  334 &  668 & 874 & 1337 \\
      & 032 & 2008-07-19 15:28 & 2008-07-19 17:28 & 54666.69 &   2140 &  176 &  176 &  176 &  352 & 443 & 704 \\
      & 033 & 2008-08-16 02:28 & 2008-08-16 05:57 & 54694.18 &   1854 &  153 &  153 &  153 &  306 & 383 & 614 \\
      & 034 & 2008-09-21 09:11 & 2008-09-21 12:27 & 54730.45 &    546 & \nodata & 36 & 114 &  346 & \nodata & 21 \\
      & 035 & 2008-10-03 13:41 & 2008-10-03 18:41 & 54742.67 &   2046 &  214 &  214 &  214 &  428 & 350 & 653 \\
      & 036 & 2008-10-17 16:26 & 2008-10-17 16:34 & 54756.69 &    452 &   41 &   41 &   41 &   83 &  74 & 165 \\
      & 037 & 2008-10-21 05:22 & 2008-10-21 07:21 & 54760.27 &   2721 &  230 &  230 &  230 &  460 & 587 & 921 \\
\enddata

\tablenotetext{1}{Observing times prior to January 2008 can be found in \citet{grupe07a} and \citet{grupe08b}.}
\tablenotetext{2}{Start and end times are given in UT}
\tablenotetext{3}{Observing time given in s}
\end{deluxetable}

\begin{deluxetable}{cccccccccc}
\tabletypesize{\tiny}
\tablecaption{\swift~ UVOT Magnitudes of \wpvs
\label{swift_results}}
\tablewidth{0pt}
\tablehead{
\colhead{ObsID}
& \colhead{Segment}
& \colhead{MJD}
& \colhead{V} 
& \colhead{B} 
& \colhead{U}  
& \colhead{UVW1} 
& \colhead{UVM2} 
& \colhead{UVW2}  
& \colhead{\auv}
} 
\startdata
30334 & 001 & 53663.15 & \nodata        &  \nodata       & \nodata        &  14.11\plm0.03 &  14.11\plm0.03 &  14.21\plm0.03 & \nodata  \\
      & 002 & 53711.57 & \nodata        &  \nodata       & \nodata        &  14.19\plm0.03 &  14.16\plm0.03 &  14.28\plm0.03 & \nodata  \\
      & 003 & 53740.16 & \nodata        &  \nodata       & \nodata        &  14.11\plm0.03 &  14.10\plm0.03 &  14.25\plm0.03 & \nodata  \\
      & 004 & 53947.23 & 15.19\plm0.04  &  15.49\plm0.03 &  14.44\plm0.03 &  14.32\plm0.03 &  14.30\plm0.04 &  14.51\plm0.02 & 0.95\plm0.11  \\
      & 005 & 53984.42 & 15.08\plm0.05  &  15.38\plm0.02 &  14.29\plm0.02 &  14.14\plm0.02 &  14.06\plm0.03 &  14.23\plm0.03 & 0.78\plm0.09  \\
      & 006 & 53989.28 & 15.11\plm0.03  &  15.39\plm0.03 &  14.31\plm0.03 &  14.13\plm0.03 &  14.08\plm0.03 &  14.22\plm0.03 & 0.77\plm0.08  \\
      & 007 & 54035.54 & 15.09\plm0.03  &  15.38\plm0.02 &  14.32\plm0.03 &  14.14\plm0.03 &  14.12\plm0.03 &  14.21\plm0.03 & 0.80\plm0.07  \\
      & 008 & 54075.14 & 15.17\plm0.03  &  15.49\plm0.03 &  14.42\plm0.03 &  14.32\plm0.03 &  14.31\plm0.04 &  14.42\plm0.03 & 0.92\plm0.08  \\
\enddata														       
\tablenotetext{1}{Magnitude corrected for reddening with $E_{\rm B-V}$=0.037 given by
\citet{sfd98}. The errors given in this table are statistical errors}
\end{deluxetable}

\begin{deluxetable}{cccc}
\tabletypesize{\tiny}
\tablecaption{\swift~ XRT 3$\sigma$ upper limits of \wpvs
\label{swift_xrt_3sul}}
\tablewidth{0pt}
\tablehead{
\colhead{ObsID}
& \colhead{Segment}
& \colhead{MJD}
& \colhead{3$\sigma$  ul} 
} 
\startdata
30334 & 015 & 54299.81 & 4.54 \\
      & 018 & 54349.96 & 2.42 \\
      & 021 & 54354.52 & 2.25 \\
      & 022 & 54355.53 & 1.82 \\
      & 023 & 54356.52 & 1.68 \\
      & 024 & 54357.52 & 1.52 \\
      & 026 & 54380.30 & 3.43 \\
      & 027 & 54381.31 & 2.54 \\
\enddata														       
\tablenotetext{1}{The 3$\sigma$ upper limits are calculated using the 
method by \citet{kraft91} and are given in units of $10^{-3}$ counts s$^{-1}$.
}
\end{deluxetable}

\begin{deluxetable}{lrccccc}
\tabletypesize{\scriptsize}
\tablecaption{Detection of WPVS 007 by \swift\
\label{xrt_stat}}
\tablewidth{0pt}
\tablehead{
\colhead{Obs date} 
& \colhead{$T_{\rm exp}$\tablenotemark{1}}
& \colhead{Source counts\tablenotemark{2}}
& \colhead{CR\tablenotemark{3}} 
& \colhead{HR\tablenotemark{4}} 
& \colhead{0.2-2.0 keV flux\tablenotemark{5}}
& \colhead{$\alpha_{\rm ox}$\tablenotemark{6}}
} 
\startdata
2007 September & 50531 & 30.0$^{+6.0}_{-5.2}$ & 5.9$^{+1.2}_{-1.0}$ & +0.54$^{+0.18}_{-0.12}$ &  1.48\plm0.33 & --- \\
2008 December 21 & 2330  & 9.7$^{+3.5}_{-2.9}$ & 41.2$^{+15.0}_{-12.5}$ & --- & 17.55$^{+6.39}_{-5.33}$\tablenotemark{7} & 1.95 \\
2009 September 17 & 1818 & 33.3$^{+6.3}_{-5.5}$ & 183.2$^{+34.6}_{-30.3}$ & $-$0.54$^{+0.13}_{-0.16}$ &  160.0$^{+30.2}_{-26.5}$ & 1.89 \\
2010 June 15 & 5020 & 6.2$^{+3.0}_{-2.3}$ & 12.4$^{+6.0}_{-4.6}$ & --- & 5.28$^{+2.56}_{-1.96}$\tablenotemark{7} & 2.38 \\
2010 June 22 & 4848 & 6.3$^{+3.1}_{-2.3}$ & 13.0$^{+6.4}_{-4.7}$ & --- & 5.54$^{+2.73}_{-2.00}$\tablenotemark{7} & 2.36 \\
2010 July 06 & 5218 & 17.9$^{+4.8}_{-4.1}$ & 34.3$^{+9.2}_{-7.9}$ & $-$0.48$^{+0.20}_{-0.25}$ & 22.76$^{+6.12}_{-5.24}$\tablenotemark{7} & 2.36 \\
2010 July 13 & 4111 & 40.6$^{+6.8}_{-6.1}$ & 98.8$^{+16.5}_{-14.8}$ & $-$0.11$^{+0.14}_{-0.16}$ & 53.4$^{+8.9}_{-8.0}$ & 1.87 \\
2010 September 21 & 3516 & 17.2$^{4.6}_{-3.9}$ & 48.9$^{+13.1}_{-11.1}$ & $-$0.25$^{+0.11}_{-0.15}$ & 22.66$^{+6.07}_{-5.14}$\tablenotemark{7} & 2.28\\
\enddata

\tablenotetext{1}{Exposure times are given in s}
\tablenotetext{2}{Source counts are background corrected and the uncertainties were determined using 
Bayesian statistics as described in \citet{kraft91}.}
\tablenotetext{3}{Count rate is given in units of $10^{-4}$ counts s$^{-1}$.}
\tablenotetext{4}{The hardness ratio is defined as HR=(H-S)/(H+S) with S and H are
background subtracted counts in the 0.3-1.0 and 1.0-10.0 keV bands, respectively. The
hardness ratio was calculated following the description in \citet{park06}. }
\tablenotetext{5}{The rest-frame 0.2-2.0 keV flux is given in units of $10^{-17}$ W m$^{-2}$,  
corrected for Galactic absorption}
\tablenotetext{6}{Observed \aox; the uncertainties in \aox\ are of the order of 0.2.}
\tablenotetext{7}{Assuming the X-ray spectrum measured during the 2010 July 13 \swift\ observation.}
\tablenotetext{8}{Assuming the mean X-ray spectrum of the 2011 September \swift\ observations.}
\tablenotetext{9}{Mean values of all data when \wpvs\ was detected during the 2011 September intensive monitoring
campaign.}
\end{deluxetable}

\begin{deluxetable}{lrcccc}
\tabletypesize{\scriptsize}
\tablecaption{Spectral analysis of WPVS 007 during detections\tablenotemark{1}
\label{xrt_spec}}
\tablewidth{0pt}
\tablehead{
\colhead{Obs date} 
& \colhead{Model\tablenotemark{2}}
& \colhead{\ax}
& \colhead{$N_{\rm H, pc}$\tablenotemark{3}} 
& \colhead{$f_{\rm pc}$\tablenotemark{4}} 
& \colhead{C-stat }
} 
\startdata
               & powl  & -0.84$^{+0.57}_{-0.88}$ & --- & --- & 50.3/40 \\
\rb{2007 September} & zpcfabs * po & 0.77$^{+0.44}_{-0.51}$ & 13.73$^{+13.86}_{-6.87}$ & 0.95 & 42.6/39 \\
               & powl  & 2.67$^{+0.76}_{-0.65}$ &  --- & --- & 20.6/26 \\
\rb{2009 September 17} & zpcfabs & 2.65$^{+1.00}_{-0.67}$ & 17.39$^{+137.83}_{-8.58}$ &  0.95 & 17.6/25 \\
               & powl  &  1.56$^{+0.47}_{-0.45}$ & --- & --- & 36.1/32 \\
\rb{2010 July 13} & zpcfabs & 2.76$^{+0.50}_{-1.34}$ & 2.80$^{+62.50}_{-1.37}$ & 0.95 & 32.1/31 \\
               & powl  &  1.66$^{+0.70}_{-0.61}$ & --- & --- & 34.4/20 \\
\rb{2011 July 23} & zpcfsabs & 2.83$^{+0.54}_{-0.56}$ & 3.02$^{+3.88}_{-1.40}$ & 0.95 & 24.1/19 \\
\enddata

\tablenotetext{1}{All spectra were fitted with the z=0 absorber fixed to the Galactic value of $N_{\rm H} = 1.4\times 10^{20}$
cm$^{-2}$.}
\tablenotetext{2}{powl = single power law model, zpcfabs * powl = partial covering absorber with single power low model}
\tablenotetext{3}{The partial covering absorber column density $N_{\rm H, pc}$ is given in units of $10^{22}$ cm$^{-2}$}
\tablenotetext{4}{Covering fraction of the partial covering absorber}
\tablenotetext{5}{Mean spectrum of all data when \wpvs\ was detected during the 2011 September intensive monitoring
campaign.}
\tablenotetext{6}{Mean spectrum of all data when WPVS 007 was not detected between October 2005 and
June 2013 as listed in Table\,\ref{swift_xrt_3sul}.}
\end{deluxetable}

\begin{deluxetable}{lccc}
\tabletypesize{\scriptsize}
\tablecaption{Infrared Observations of \wpvs\ with 2MASS and WISE.
\label{wpvs_ir}}
\tablewidth{0pt}
\tablehead{
\colhead{Filter} 
& \colhead{$\nu_{\rm c}$\tablenotemark{1}}
& \colhead{Flux\tablenotemark{2}}
& \colhead{$\nu L_{\nu }$\tablenotemark{3}} 
} 
\startdata
2MASS J & 2.40 & 1.7568\plm0.10656 & 2.94 \\
2MASS H & 1.82 & 1.4888\plm0.12867 & 2.49 \\
2MASS K & 1.38 & 1.5456\plm0.095634 & 2.59 \\
WISE 3.4$\mu$ & 0.8856 & 1.5428\plm0.030132 & 2.58 \\
WISE 4.6$\mu$ & 0.6445 & 1.4232\plm0.025126 & 2.38 \\ 
WISE 12$\mu$ &  0.2675 & 1.2072\plm0.016795 & 2.02 \\
WISE 22$\mu$ &  0.1346 & 1.2472\plm0.029052 & 2.09 \\
\enddata

\tablenotetext{1}{Central filter frequency $\nu_{\rm c}$ in unites of $10^{14}$ Hz}
\tablenotetext{2}{Flux $\nu F_{\nu}$ and flux uncertainty are given in units of  $10^{-14}$ W m$^{-2}$}
\tablenotetext{3}{The luminosity $\nu L_{\nu}$ is given in units of $10^{36}$ W.}
\end{deluxetable}

\begin{deluxetable}{lccc}
\tabletypesize{\scriptsize}
\tablecaption{NIR emission lines in WPVS 007.
\label{nir_lines}}
\tablewidth{0pt}
\tablehead{
\colhead{Line} 
& \colhead{$\lambda_0$\tablenotemark{1}}
& \colhead{FWHM\tablenotemark{2}} 
& \colhead{Flux\tablenotemark{3}}
} 
\startdata
Mg II & 9218+9244 & --- & 43.8 \\
FeII + Pa$\epsilon$ & 9549 + 9573 & --- & 20.7 \\
Fe II & 9997 & 1680 & 15.5\plm3.8 \\
Pa$\delta$ & 10052 & 1310 & 16.5\plm3.4 \\
Fe II & 10501 & 1570 & 10.6\plm0.9 \\
He I & 10833 & 1490 & 43.3\plm4.5 \\
Pa$\gamma$ & 10941 & 1730 & 34.7\plm5.0 \\
Fe II & 11126 & 1070 & 4.7\plm0.9 \\
O I & 11290 & 1320 & 18.0\plm1.7 \\
O III & 12470 & 990 & 5.3\plm0.7 \\
Fe II & 12570 & 1300 & 4.2\plm0.9 \\
Pa$\beta$ & 12822 & 1230 & 39.3\plm3.2 \\
O III & 13436 & 885 & 13.8\plm2.0 \\
Fe I & 14298 & --- & --- \\
Fe I & 14480 & --- & --- \\
Fe I & 14628 & --- & --- \\
Fe I & 15260 & --- & --- \\
Fe I & 15335 & --- & --- \\
Fe I & 15463 & --- & --- \\
Fe I & 15533 & --- & --- \\
Fe I & 15680 & --- & --- \\
Fe I & 15956 & --- & --- \\
Fe I & 16366 & --- & --- \\
Fe I & 16445 & --- & --- \\
Fe I & 16885 & --- & --- \\
Fe I & 17383 & --- & --- \\
Fe I & 17675 & --- & --- \\
Pa$\alpha$ & 18751 & 1035 & 53.4\plm5.6 \\
He I & 18922 & 1400 & 11.0\plm1.3 \\
Br $\delta$ & 19400 & --- & --- \\
Mg II & 21064 & 775 & 4.9\plm0.6 \\
Br $\gamma$ & 21661 & 1130 & 6.5\plm0.7
\enddata

\tablenotetext{1}{Rest wavelength $\lambda_0$ given in units of \AA}
\tablenotetext{2}{Full Width at Half Maximum (FWHM) given in units of km s$^{-1}$. Note that for the FeI lines FWHM and flux can not be given 
reliably because these lines appear to be blended and often so faint that they are affected by noise.}
\tablenotetext{3}{The Line flux $\nu F_{\nu}$ and flux uncertainty given in units of  $10^{-18}$ W m$^{-2}$ 
(10$^{-15}$ erg s$^{-1}$ cm$^{-2}$).}
\end{deluxetable}

\end{document}